# Probing plexciton dynamics with higher-order spectroscopy

Simon Büttner[1#], Luca Nils Philipp[1#], Julian Lüttig[1,2], Maximilian Rödel[3], Matthias Hensen[1], Jens Pflaum[3,4‡], Roland Mitric[1+], and Tobias Brixner[1,5*]

*[1]Institut für Physikalische und Theoretische Chemie, Universität Würzburg, Am Hubland, 97074 Würzburg, Germany*
*[2]Department of Physics, University of Michigan, 450 Church Street, Ann Arbor, Michigan 48109, USA*
*[3]Experimental Physics VI, University of Würzburg, Am Hubland, 97074 Würzburg, Germany*
*[4]Center for Applied Energy Research e.V. (CAE Bayern), Magdalene-Schoch-Straße 3, 97074 Würzburg, Germany*
*[5]Center for Nanosystems Chemistry (CNC), Universität Würzburg, Theodor-Boveri-Weg, 97074 Würzburg, Germany*

*[#]Contributed equally to this work*

*[‡]Electronic mail: jens.pflaum@uni-wuerzburg.de*
*[+]Electronic mail: roland.mitric@uni-wuerzburg.de*
**Electronic mail: tobias.brixner@uni-wuerzburg.de*

## Abstract

Coupling molecular transition dipole moments to surface-plasmon polaritons (SPPs) results in the formation of new optical quasiparticles, i.e., plexcitons. Mixing the specific properties of matter excitations and light modes has proven to be an efficient strategy to alter a variety of molecular processes ranging from chemical reactions to exciton transport. Here, we investigate energy transfer in a plexcitonic system of zinc phthalocyanine (ZnPc) molecules aggregated in the crystalline $\alpha$-phase and an SPP on a planar gold surface. By tuning the angle of incidence, we vary the degree of mixing between excitonic and SPP character of the excited state. We apply our recently developed higher-order pump–probe spectroscopy to separate the system's fifth-order signal describing the dynamics of two-particle interactions. The time it takes for two quasiparticles to meet and annihilate is a measure of their movement and thus the transport of excitation energy in the system. We find that the transport extracted from the fifth-order signal is surprisingly unaffected by the mixing ratio of exciton and SPP contributions of the plexciton. Using a rate equation model, we explain this behavior by fast transition from the plexcitonic states to many localized excitonic dark states that do not have an SPP contribution. Our results give an indication of how hybrid exciton–plasmon systems should be designed to exploit the delocalization of the involved plasmon modes for improved transport.

# I. Introduction

Energy transport plays a crucial role in light harvesting and solar energy conversion. The improvement of transport properties using, e.g., coupled molecular systems is a widely pursued path towards the cost-effective production of highly efficient optoelectronic devices. Using molecules as entities for transporting excitation energy offers chemists and material scientists enormous flexibility in preparing new materials, shaping their structural composition and, thus, enhancing their technological potential.[1–6] In addition, especially in recent years, an approach has been pursued in which the molecular quasiparticle carrying the excitation energy, i.e., the exciton, is converted into a hybrid quasiparticle which by nature has the character of matter and light, the so-called molecular polariton. This optical quasiparticle is created by coupling of molecular excitons to specifically tailored modes of the electromagnetic field. In particular, the delocalization of these modes is a crucial aspect in the improvement of excitonic transport properties, as this is intended to increase the transport range that is usually determined by incoherent hopping within the organic semiconductor. In addition, hybrid excitations might reduce loss channels based on the recombination of electrons and holes.[7–10]

While in this paper we show a molecular polariton which emerges by strongly coupling to an SPP, a large number of studies investigated molecular polaritons which are formed by placing molecules inside an optical microcavity whose resonance frequency is tuned to the exciton. Subsequently, the interaction between the excited electronic states of the molecules and the electromagnetic field modes of the optical microcavity leads to the formation of new hybrid eigenstates.[11–13] The formation of hybrid light–matter states alters various photophysical properties. In one example, this was achieved by bringing one of the main antenna complexes of LH2 into a cavity, resulting in a prolongation of the excited-state lifetime compared with the bare exciton.[14,15] In another recent example, two thin layers of J-aggregated donor and acceptor molecules separated by a spacer of 2 μm were inserted into a microcavity and the resulting polariton states were delocalized over the whole system.[16] For an ensemble of $N$ molecules and a single cavity mode the single-particle manifold contains the so-called lower polariton state (LP), the upper polariton state (UP), and an ensemble of $N-1$ dark states (DS) that are energetically positioned in between LP and UP. In linear absorption, the LP and UP appear as two distinct peaks, separated by the so-called Rabi splitting, whereas transitions from the ground state (GS) to any of the DS are forbidden such that the dark states cannot be observed in absorption experiments.

In contrast to the delocalized LP and UP, the DS are purely molecular states located at the energy of the molecular exciton although they result from the coupling of many molecules with a common light mode. Thus, most properties of the DS are expected to closely resemble those of the molecular system uncoupled to the electromagnetic field mode. In this way, energy can be transferred from the LP and UP to the DS, which leads to the observation of molecular-like properties for polaritonic samples.[16] Even though such effects have been observed, a generally accepted microscopic description of the relaxation from the bright polaritonic states to the DS is yet to be found. While recent theoretical studies suggest that the relaxation process is closely connected with cavity losses,[17,18] others have suggested that the transfer from the LP to the DS (as well as the UP to DS) is caused by phonon-assisted scattering.[19,20]

Besides microcavities, plasmonic systems, which are the subject of this work, are particularly suitable for locally modifying the transport of excitons in molecular layers. In these plasmonic systems, molecular excitons are resonantly coupled to the electromagnetic near-field of, e.g., surface-plasmon polariton (SPP) modes that are strongly confined to the molecule–metal interface but can be strongly delocalized along the lateral dimensions. The quasiparticle created by the coupling is referred to as a plexciton. For example, Berghuis and coworkers were able to show an example in which the propagation length of plexcitons increased by two orders of magnitude compared to the hopping-type propagation of pure excitons.[10] In their work, the strong coupling regime was reached by exposing excitons of a tetracene crystal to surface lattice resonances formed by a plasmonic nanoparticle array. Other studies investigated the enhancement of delocalization by coupling excitons of a squaraine dye aggregate to SPP modes of a plain gold surface,[21,22] modification of coherences of nanohybrid systems made from J-aggregates and gold nanourchins,[23] or the various relaxation pathways of J-aggregates to gold nanowires.[24] A significant difference between polaritons emerging from a cavity and those from an SPP lies in the distribution of electromagnetic field modes. In a cavity, discrete light modes contribute to the formation of polaritons. Conversely, SPPs provide a range of continuous light modes to create molecular polaritons.

Linear spectroscopy is a standard method for characterizing the properties of polaritonic systems. For instance, absorption spectroscopy is employed for measuring the coupling strength that is related to the Rabi splitting of the emerging polaritonic states in hybrid materials.[25] However, to probe the ultrafast dynamics of a system, nonlinear spectroscopy such as pump–probe (PP) techniques are necessary. Nonlinear spectroscopy techniques have been employed extensively to plexcitonic systems to investigate their dynamics.[26–29,25,30] Pump–probe spectroscopy is a well-suited tool to investigate energy transfer in systems where the transfer is connected to a change in the absorption spectrum. This principle can be applied in few-chromophore systems whose spectral features are clearly separated, e.g., when observing energy transfer in light-harvesting complexes with spectrally separated chromophores[31].

However, when the number of chromophores is too large, they cannot be spectrally separated anymore. This is generally the case, e.g., for polymers, polycrystalline thin films, or molecular crystals. While the excitonic bands still have a certain spectral width, spectral features cannot be assigned to specific spatial positions or chromophores. While homogeneous and inhomogeneous broadening are present in the absorption spectrum, i.e., there is a static and dynamic distribution of site energies which is random, on time average the exciton states are isoenergetic as a function of space. Under such common circumstances of an isoenergetic landscape of excited states, the absorption spectrum does not change when the exciton is transported because the average excitation energy does not depend on position, and hence it does not change when the exciton moves. Consequently, normal time-resolved spectroscopy that is of third order in nonlinear response does not reveal signatures of energy transport. This limitation does not only apply to PP but also to more advanced third-order techniques such as coherent two-dimensional electronic spectroscopy that is very well suited to reveal energy transfer in light-harvesting systems.[32,33] For spectrally resolvable systems, we have proven mathematically that the state-to-state population transfer rates can be recovered uniquely.[34] For larger systems with spectrally unresolved states, different methods have to be employed.

In literature, transport has been characterized with combined spatial and temporal resolution such as transient absorption laterally resolved by microscopy[35] or the time-resolved imaging of photoluminescence.[36–38] Another method employs the interaction of excitons, i.e., exciton–exciton annihilation, to probe exciton transport in isoenergetic systems together with suitable modeling of exciton movement.[39,40,22,41] Exciton interaction can only occur if the excitons are in close proximity to each other. The time required for the excitons to meet and its dependence on the quasiparticle density, in combination with a diffusion model, enables the quantification of the propagation length of the excitons.

Typically, such measurements are conducted by performing PP spectroscopy at different pump-pulse intensities. For low pump intensities the third-order signal dominates, while for higher pump intensities also higher orders of nonlinearity contribute to the signal. While the third-order signal describes single-exciton dynamics, the interaction of two excitons is encoded in the fifth-order signal. Modeling the intensity-dependent data allows one to extract the exciton diffusion properties. However, in such intensity-dependent measurements, the third- and fifth-order responses, i.e., the single- and two-quasiparticle dynamics, are mixed and difficult to separate without making a-priori assumptions. Furthermore, higher-order contributions, i.e., due to the interaction of three, four, and more quasiparticles, might also contribute. We have recently demonstrated a technique that can disentangle the different orders of nonlinearity from intensity-dependent PP spectra.[42–44] This allows one to separate the pure third- and fifth-order responses, independent of any assumptions about the underlying quantum system. In the case of an excitonic system, one can then identify the third-order signal with the single-exciton kinetics and obtain the clean two-exciton kinetics from the separated third- and fifth-order signals. This allows one to quantify exciton diffusion even for on-average isoenergetic many-chromophore systems.

Here, we employ higher-order PP spectroscopy on a plexcitonic system. Our sample consists of a polycrystalline zinc phthalocyanine (ZnPc) thin film on gold layer on glass (ZnPc/Au). We used ZnPc because it is photostable and because it has well-known optical properties in thin films.[45,46] We measured the ZnPc/Au sample in an experimental configuration that allows us to tune the degree of plasmonic versus excitonic contributions to the coupled plexcitonic regime by varying the incidence angle in Kretschmann geometry. We separate the ultrafast third- and fifth-order responses as a function of plexcitonic character. By comparing the plexcitonic with the bare excitonic dynamics we disclose the role of coupling for the overall system dynamics and transport properties.

## II. Method and Experimental Setup

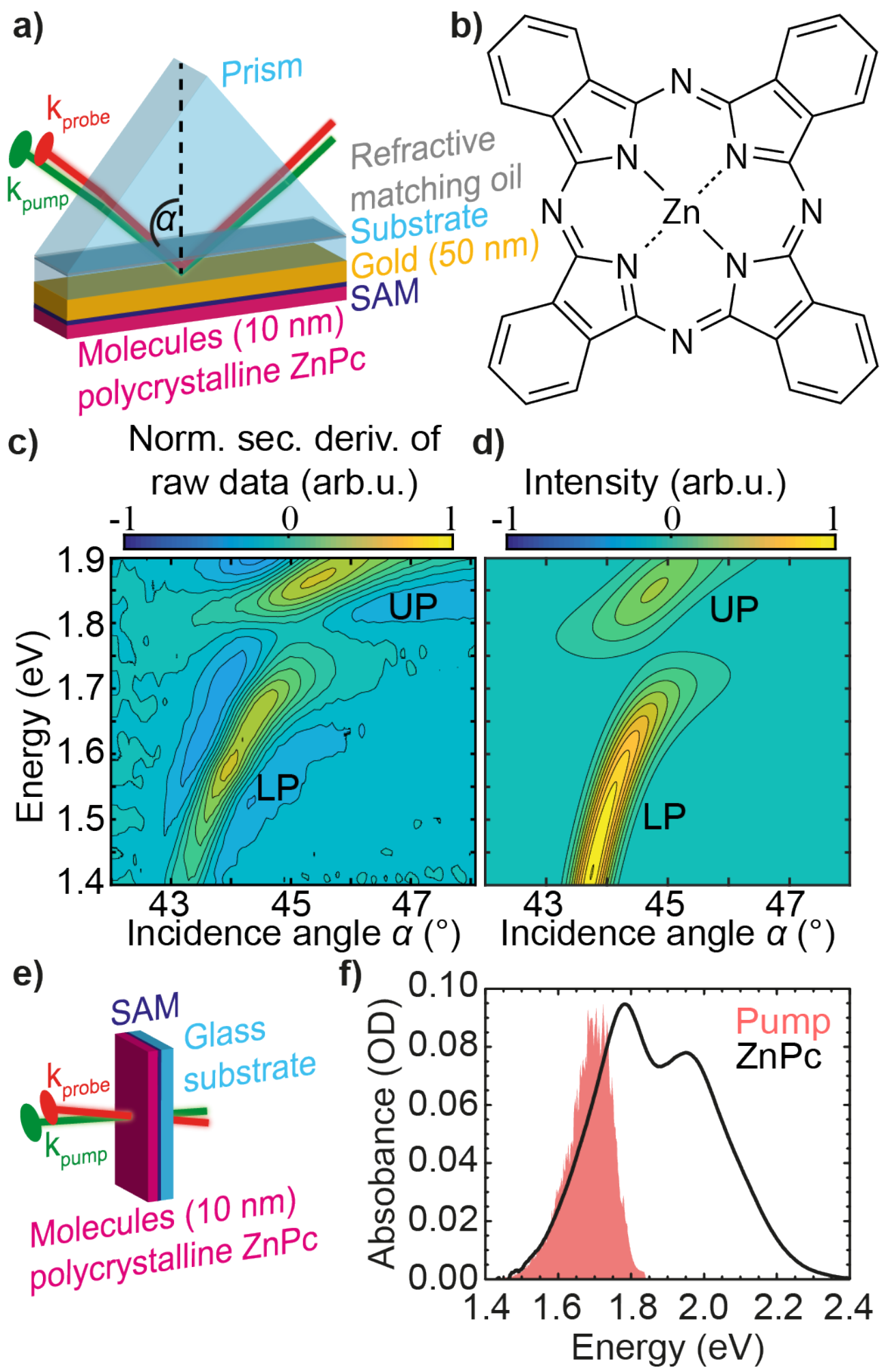

Figure 1. Experimental configurations for the plexcitonic and purely excitonic sample. Plexciton sample: (a) Schematic measurement geometry of polycrystalline zinc phthalocyanine (ZnPc) thin film on a self-assembled monolayer (SAM) functionalized gold layer on a glass substrate. Pump and probe beams are coupled onto the sample via a prism and hit the gold surface under the same incidence angle α. (b) Chemical structure of ZnPc. (c) Normalized second derivative of measured difference reflection spectra. (d) Plexciton dispersion spectra calculated within the Tavis–Cummings model. Excitonic sample: (e) Schematic measurement geometry of purely excitonic crystalline ZnPc thin film. (f) Absorption of the bare ZnPc film on glass (black) and the pump laser spectrum (red).

Our coupled plexcitonic sample, labeled ZnPc/Au (Fig. 1a), consists of a 10 nm thick polycrystalline ZnPc (Fig. 1b) film and a gold layer of 50 nm thickness evaporated on a glass substrate. To prevent quenching of the excited quasiparticles at the metal-organic interface, a self-assembled monolayer (SAM) of 1-decanethiol was placed between the ZnPc film and the gold surface (for a detailed description of the sample preparation see Supplementary Information (SI), Section I). To excite SPPs at the ZnPc/Au boundary, a prism was positioned on the glass substrate with a reflective matching oil in between. In this geometry, the laser beam hits the sample at an incidence angle $\alpha$ from the glass side (Fig. 1a). By varying $\alpha$, SPPs with different $k$-vector and energies are excited according to the underlying dispersion. Tuning the SPP resonance to that of the excitonic system usually leads to the archetypal avoided crossing in the dispersion relation of the investigated system. We have experimentally verified this behavior for our sample and show the measured dispersion relation in Figure 1c. To emphasize the splitting of LP and UP more clearly, we took the second derivate along the energy axis of the underlying difference reflection spectrum. Changing the energy of the SPP via the angle of incidence $\alpha$ allows us to control the ratio of the excitonic and the plasmonic contribution: A small angle $\alpha$ results in a small in-plane wave vector of the incident field and SPPs with a long wavelength, or small excitation energy, are excited. If the SPP excitation energy is below the exciton energy of the aggregated ZnPc, the coupling-induced LP state exhibits mainly SPP character. If the angle $\alpha$ is increased, the SSP character of the LP becomes less prominent and the exciton character increases. At the incidence angle $\alpha$ for which the SPP excitation energy matches the exciton energy, the resulting LP has equal proportions of excitonic and plasmonic character.[27] This opens the possibility for the investigation of the transport and annihilation dynamics of the same sample at different plexciton compositions. Note that the ratio of SPP and exciton in the UP is opposite to that in the LP.

The measured reflection spectra are due to the dispersion relation described by the Tavis–Cummings model (Fig. 1d). The ZnPc/Au sample and the reflection spectra are known from literature.[47] While in literature so called quasi-dark ("gray") states are reported,[48] we cannot observe such states due to the linewidth of the plexcitonic states and the low dipole moments of these quasi-dark states. Therefore, we neglected this effect also in the simulations. To compare the plexcitonic system with a purely molecular excitonic system, we also measured a sample without Au thin film (ZnPc/glass, Fig. 1e). The time-depended measurements in this case were performed under transient absorption (TA) geometry with the pump laser spectrum and the static absorption spectrum of the ZnPc/glass sample shown in Figure 1f.

For the time-resolved measurements, we used higher-order PP spectroscopy, which is a nonlinear optical technique that enables the separation of single-particle dynamics from multi-particle interactions. In conventional PP measurements contributions from various nonlinear orders are mixed, making it challenging to distinguish between different excitation phenomena. The higher-order PP spectroscopy method uses intensity-dependent measurements to separate signals of different nonlinear order systematically.[42–44]

In general, the PP signal of a system can be expanded in a perturbative series based on the excitation (pump) intensity $I_p$,

$$\mathrm{PP}(I_p) = \sum_{r=1}^{\infty} \mathrm{PP}^{(2r+1)} I_p^r, \tag{1}$$

where $\mathrm{PP}^{(2r+1)}$ represents the $(2r+1)^{\text{th}}$-order nonlinear response with respect to the electric field. To isolate different nonlinear orders, we use "intensity cycling" of the pump pulse. A specific number $M$ of PP signals are measured with a varied pump intensity $I_p$,

$$I_p = 4I_0 \cos^2\left(\frac{(p-1)\pi}{2M}\right), p = 1, \dots, M, \tag{2}$$

for each measurement $p$, where $I_0$ is a "base" pump intensity. The highest order which can be corrected (and separated) in this scheme is the $(2M+1)^{\text{th}}$ order and therefore depends on the number of "intensity-cycling" steps $M$. The specific choice of pump intensities $I_p$ in Eq. 2 arises in a derivation that considers so-called $n$-quantum signals $\mathrm{PP}^{(n\mathrm{Q})}$.[42] These $\mathrm{PP}^{(n\mathrm{Q})}$ can be extracted by a weighted summation of the recorded PP signals at different $I_p$ values,

$$\mathrm{PP}^{(nQ)}(I_0) = \sum_{p=1}^{M} w_p^{(n\mathrm{Q})} \mathrm{PP}(I_p), \tag{3}$$

where the weights $w_p^{(n\mathrm{Q})}$ are given as

$$w_p^{n\mathrm{Q}} = \frac{1}{2M} \frac{2-\delta_{p,1}}{1+\delta_{n,M}} \cos\left(\frac{n(p-1)2\pi}{2M}\right), \tag{4}$$

where $\delta_{j,k}$ is the Kronecker delta. Each of these $n$-quantum signals $\mathrm{PP}^{(n\mathrm{Q})}$ is a linear combination of different $(2r+1)^{\text{th}}$-order nonlinear responses. To isolate the $\mathrm{PP}^{(2r+1)}$ signals, the equation

$$\mathrm{PP}^{(n\mathrm{Q})}(I_0) = \sum_{r=n}^{M} \binom{2r}{r-n} \mathrm{PP}^{(2r+1)} I_0^r = \sum_{r=n}^{M} \Lambda_r^{n\mathrm{Q}} \mathrm{PP}^{(2r+1)} I_0^r \tag{5}$$

can be used, where $\binom{a}{b}$ denotes the binomial coefficient. The coefficients $\Lambda_r^{n\mathrm{Q}}$ form a matrix $\Lambda$ that can be inverted by $\Lambda^{-1} = \sum_{a=0}^{M-1} (\mathbb{I} - \Lambda)^a$, where $\mathbb{I}$ is the identity matrix, finally leading to

$$\mathrm{PP}^{(2r+1)} I_0^r = \sum_{n=r}^{M} (\Lambda^{-1})_{n\mathrm{Q}}^{r} \mathrm{PP}^{(n\mathrm{Q})}(I_0). \tag{6}$$

This method allows the isolation of contributions from different nonlinear orders while eliminating contamination from higher-order effects. For example, to get a pure third-order signal $\mathrm{PP}^{(3)}$ without fifth- and seventh-order contaminations, a total of $M = 3$ PP measurements are needed. In this case, the pump intensities $I_p$ are $I_1 = I_0$, $I_2 = 3I_0$ and $I_3 = 4I_0$, according to Eq. 2, and Eq. 6 can be simplified to

$$\mathrm{PP}^{(3)}(I_0) = 2\mathrm{PP}(I_0) - \tfrac{2}{3}\mathrm{PP}(3I_0) + \tfrac{1}{4}\mathrm{PP}(4I_0). \quad (7)$$

The higher-order signals $\mathrm{PP}^{(5)}$ and $\mathrm{PP}^{(7)}$ can be extracted in a similar way with the same recorded pump-intensity-dependent PP spectra $\mathrm{PP}(I_p)$ as in Eq. 7 via

$$\mathrm{PP}^{(5)}(I_0) = -\tfrac{7}{6}\mathrm{PP}(I_0) + \tfrac{5}{6}\mathrm{PP}(3I_0) - \tfrac{1}{3}\mathrm{PP}(4I_0) \quad (8)$$

and

$$\mathrm{PP}^{(7)}(I_0) = +\tfrac{1}{6}\mathrm{PP}(I_0) - \tfrac{1}{6}\mathrm{PP}(3I_0) + \tfrac{1}{12}\mathrm{PP}(4I_0). \quad (9)$$

A more detailed description can be found in the literature.[42,44]

This approach allows one to systematically increase the number of probed quasiparticle interactions. For example, the third-order signal ($\mathrm{PP}^{(3)}$) reflects single-particle dynamics, while higher orders ($\mathrm{PP}^{(5)}$, $\mathrm{PP}^{(7)}$, etc.) reveal interactions between multiple particles, in large systems.[42] The method enables the study of multi-particle effects, such as exciton–exciton annihilation, multi-exciton formation, or other higher-order quasiparticle interactions. From the dynamics of the higher-order signal the quasiparticle diffusion can be inferred. It can be shown that the fifth-order signal is sensitive to quasiparticle–quasiparticle annihilation and therefore sensitive to the their transport.[42,49,50] Annihilation occurs only when quasiparticles are in close proximity. When quasiparticles are randomly distributed after the initial excitation step, they must subsequently move through the material to encounter each other and generate a higher-order signal. Consequently, the time required for annihilation is dependent on the transport process of the quasiparticles and their density.

The experimental PP setup consisted of a 100 kHz Yb laser (Pharos, Light Conversion, 1030 nm, 20 W, 485 fs) whose output was guided to a noncollinear optical parametric amplifier (Orpheus, Light Conversion). The resulting broadband white light was recollimated and split into two beams with a 10:90 broadband beamsplitter (Layertec). The less intense beam was used as probe beam and was compressed with a quartz glass prism compressor, widened with a telescope, and guided to a mechanical stage (M-IMS1000LM-S, Newport) with a mounted retroreflector to scan the population time *T*. The probe beam was attenuated by using a reflection from a wedged window and an attenuation wheel. This beam was then focused with a curved Ag mirror ($f$ = 375 mm, Eksma Optics) on the sample. The more intense pulse was used as the pump beam. The pump pulse was pre-compressed by a BK7 prism compressor and the spot size reduced by a telescope. The pump beam was further directed to an acousto-optic modulator pulse shaper (AOM, QuickShape Visible, Phasetech). The AOM pulse shaper was used to block every second pump pulse for the PP measurement, to control the intensities for the intensity cycling procedure, and to further compress the pulse. The polarization of the pump beam was set to the same polarization as the probe beam by using a half-wave plate, and the pump beam was focused with a curved Ag mirror ($f$ = 500 mm, Eksma Optics) on the sample.

The PP measurement was performed in the so-called Kretschmann geometry (Fig. 1a). To minimize spatial and temporal dispersion, the prism (330-0105, BK7, Eksma) was hit close to its edge to keep the propagation length through glass as short as possible. The sample was placed on a rotation stage (M-481-A, Newport) to vary $\alpha$. The reflected probe beam was measured by a spectrometer and a CCD line scan camera (FL3030, Entwicklungsbüro Stresing). While the plexcitonic system was measured in transient reflection geometry, the measurements with the bare molecular film were carried out in transient absorption geometry by using the transmitted beam (Figs. 1a and 1e). The pulse duration of pump and probe pulses were determined via spectral phase interferometry for direct electric-field reconstruction[51] (FC Spider VIS, APE). The duration of the pump pulses was ~19 fs [intensity full width at half maximum (FWHM)] and the probe pulse duration was ~16 fs (intensity FWHM). The prism's dispersion was accounted for during pulse compression by inserting a glass plate into the pump and probe beams. The focus spot of the pump beam at the sample position was of elliptical shape with a short axis length of about 70 μm (intensity FWHM) and a long axis length of about 130 μm (FWHM). The probe beam focus spot was of circular shape with a diameter of about 35 μm (FWHM).

## III. Results and Discussion

### III.1 Higher-Order Pump–Probe (PP) Spectroscopy

Third- and fifth-order PP spectra of the ZnPc/Au sample were measured at three different incidence angles, i.e., $\alpha$= 44.1°, $\alpha$= 44.5°, and $\alpha$= 44.9° (Fig. 2a). To separate the different nonlinear orders of the system dynamics we conducted measurements at four different pump intensities for every angle in the case of the ZnPc/Au sample and for the ZnPc/glass sample. The highest three intensities for the intensity-cycling scheme ($I_0$, $3I_0$, and $4I_0$) were chosen according to Eq. 2. With the chosen three intensities, the seventh-order signal was lower than the noise level, making the fifth-order signal the highest-order signal detectable above the noise floor. In addition, we employed a fourth pump intensity that was set lower than the other three and low enough such that higher-than-third-order contributions could be neglected. This measurement acted as a reference and was used to verify that the order extraction worked correctly by comparing the extracted third-order measurement with the low-power reference (SI, Section III, Tab. S1).[42] Combining the data obtained with these three intensities, the different nonlinear orders of the measured signal were separated via Eq. 6. In addition to the plexcitonic ZnPc/Au sample measurements, third- and fifth-order PP spectra of a ZnPc/glass sample were measured (Fig. 2b). This enables us to compare plexcitonic dynamics with pure excitonic dynamics. For the third-order PP spectra, we adopt the sign convention that ground-state bleach (GSB) and stimulated emission (SE) are treated as negative signal contributions, while excited-state absorption (ESA) is positive.

The third-order PP spectrum of the ZnPc/glass sample consists of a single negative peak (Fig. 2b), composed of GSB and SE and centered around 1.75 eV, which is known to be the main exciton transition of the ZnPc thin film (Fig. 1f). Additionally, a small positive signal is seen below 1.5 eV, which is at a lower energy than the absorption peak of the sample. In contrast to the ZnPc/glass sample, the third-order PP spectra of the ZnPc/Au sample consist of positive and negative spectral contributions above 1.5 eV for all three angles of incidence. Since we directly excite the LP, the negative signal partially arises due to GSB of the corresponding transition. Furthermore, SE will contribute to the negative signal if the system does not relax out of the LP, whereas positive signals in third-order PP spectra reflect ESA. For all three angles the positive and negative signals overlap along the energy axis so that the spectra exhibit a dispersive line shape at a specific population time $T$. Aside from the differences in the spectral shape, the spectral position of the LP GSB signal in the third-order PP spectra shows a strong angle dependence as is also observed in the linear plexciton spectra (Fig. 1c). With increasing angle of incidence, the dispersive line-shape feature shifts to higher energies in the third-order PP spectra since the SPP resonance frequency increases with increasing incidence angles. The PP spectrum at 44.1° features an additional negative spectral contribution at 1.75 eV. We attribute this to the UP. At small $\alpha$, such as $\alpha$= 44.1°, the excitation energy of the UP should be at the excitation energy of the pure exciton. However, the absolute value of the negative signal at 1.75 eV is small compared to the LP contribution and we therefore assume that it does not contribute significantly to the signal. As a result, we did not include the UP population in our simulations. At higher incidence angles the UP shifts to higher energies and cannot be excited with our laser spectrum.

For delay times larger than $T = 1$ ps, the signal dynamics in all three third-order PP spectra of the ZnPc/Au sample are similar to those of the ZnPc/glass sample, i.e., all signals decay with similar rates. This behavior was not expected, since the short lifetime of the SPP should shorten the lifetime of the plexciton compared to that of the pure exciton drastically. Upon closer inspection, the short-time evolution of some of the third-order PP spectra show some subtleties: If all third-order PP signals resulted exclusively from the decay of the LP back to the GS, one would expect an exponential decay of positive as well as of negative signals with the same rate (SI, Section IX, Fig. S4). However, the third-order PP spectrum at $\alpha$= 44.5° clearly shows an increase in the absolute value of the negative signal up to about 1 ps, while the positive signal only decays. Similar behavior is observable for the third-order PP spectrum at an incidence angle of $\alpha$= 44.9°. This implies that the LP decay dynamics do not only consist of the relaxation back to the GS, but that intermediate states are populated.

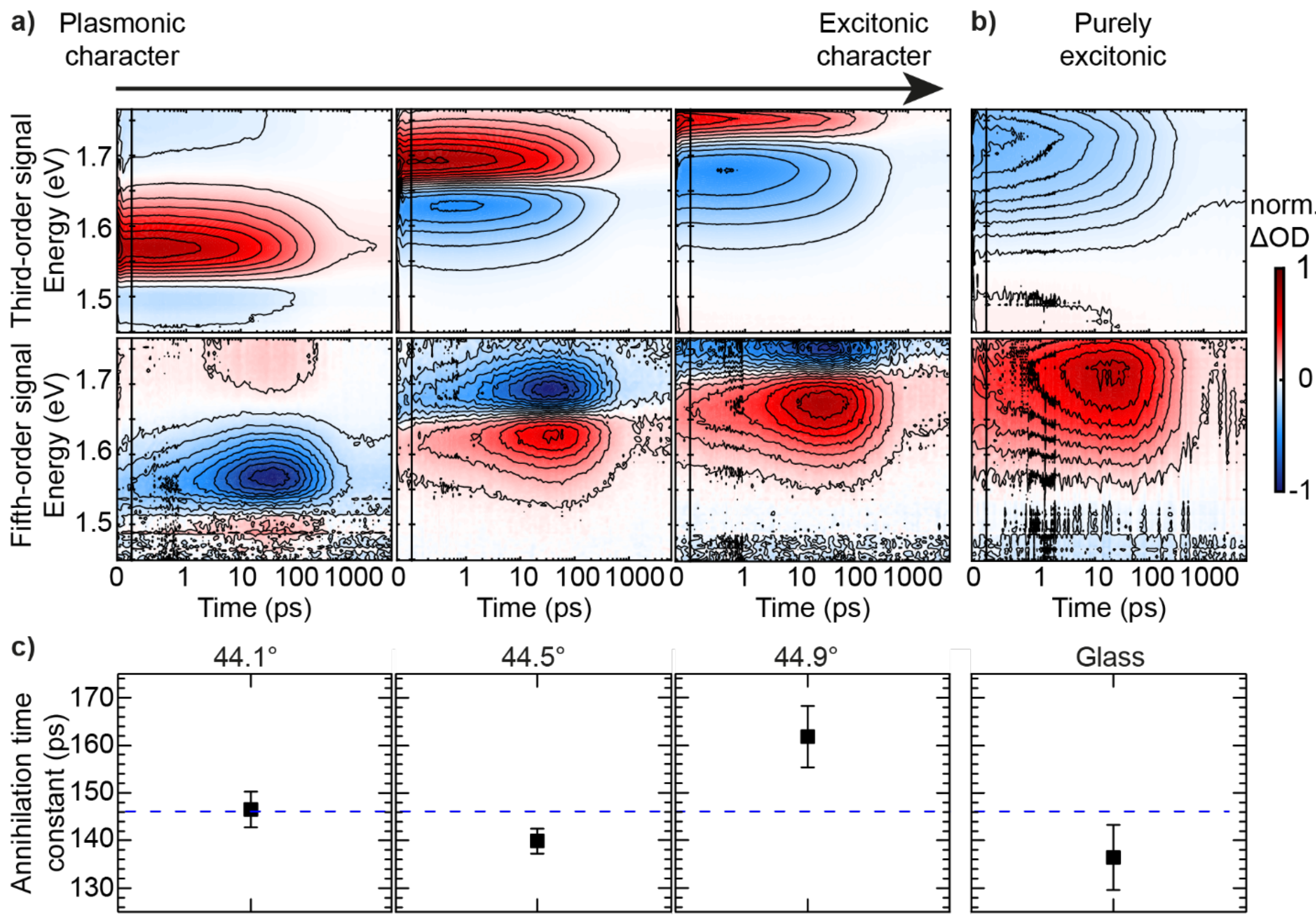

Figure 2. Time-resolved spectroscopy results for ZnPc/Au and ZnPc/glass. (a) Third-order (top row) and fifth-order (bottom row) pump–probe maps for ZnPc/Au at incidence angles of 44.1° (left), 44.5° (middle), and 44.9° (right). The excitonic character increases from left to right as indicated by the black arrow. (b) Third-order (top panel) and fifth-order (bottom panel) pump–probe maps for the purely excitonic ZnPc/glass sample. (c) Annihilation time constants extracted with global analysis. The dashed blue line indicates the mean value of all four times (146 ps).

The fifth-order PP spectrum of the ZnPc/glass sample consists of a positive contribution at the spectral position (1.75 eV) of the GSB/SE in the third-order PP data and a small positive contribution below 1.5 eV. Like the third-order PP spectra of the ZnPc/Au sample, the fifth-order PP spectra show a strong dispersive line shape of positive and negative signals above 1.5 eV. For the ZnPc/Au sample, we observe the same general trend in the fifth-order data as in the third-order data that the energy positions of the positive and negative signals depend on the incidence angle $\alpha$. However, the signal evolution does not depend on the incidence angle. In contrast to the third-order PP spectra, positive signals of fifth-order PP spectra cannot directly be uniquely attributed to ESA, and, in the same way, negative signals cannot be assigned to GSB and SE. This is because there are double-sided Feynman-diagrams for the ESA which lead to positive signals and other diagrams which lead to negative signals. The same is also true for the SE signal.[42,49,52] This complicates the interpretation of the spectral shape of fifth-order spectra. However, the temporal evolution is still indicative of excited state annihilation and thus energy transport.

We fitted the time evolution of the signal with a multi-exponential decay model in global analysis (Glotaran 1.5.1[53], the fit values are reported in SI, Section V, Tab. S2).[54] The annihilation rate is determined by the rate with which the signal of the fifth-order PP spectrum rises.[42,50,49] To obtain the annihilation rate, we carried out global analysis of the third-order signal as a first step and used the extracted third-order time constants as given quantities in the subsequent global analysis of the fifth-order signal. Note that the single-exciton-state lifetime exceeds the maximum measured population time and could not be properly resolved. We therefore set the time constant to 10 ns, based on the literature.[55] The additional time constant retrieved in the global fit of the fifth-order signal is the inverse of the annihilation rate.[42] We display in Fig. 2c the annihilation times for each measurement. We added a dashed blue line that indicates the mean value of all four annihilation times, for an easier comparison. All four times are very similar. We observe no systematic trend with increasing exitonic character (from left to right).

Consequently, the annihilation rates appear to be independent of how strong SPP and exciton contribute to the LP, and the rates themselves are almost identical to the annihilation rate of the ZnPc/glass sample. To explain the finding of the unaffected annihilation rate (and thus the plexciton transport) in our system, we simulate the early-time dynamics of the third-order spectra within a Tavis–Cummings model, to gain inside in the population kinetics of the system.

### III.2 Modeling of the Plexcitonic System

As evident from the fifth-order PP spectra, the annihilation rate in the ZnPc/Au sample is not altered by the light–matter interaction between the SPP and the main exciton transition of ZnPc compared to the pure ZnPc sample. From the unaffected annihilation rate, we conclude that the long-time energy transport does not depend on the coupling between the SPP and the exciton. Furthermore, the increase in the GSB signal of the third-order PP spectra over population time $T$ suggests that the decay of the LP does not occur solely to the GS, but that the LP also relaxes into other intermediate states with lifetimes similar to the purely excitonic states. In the present section, we substantiate this idea by carrying out simulations of the third-order PP spectra of the plexcitonic sample. For this, we propose that the LP can also relax to the DS and discuss whether this assumption is in accordance with experimental observations. This will allow us to identify certain spectral features that indicate the transition from the LP to the DS and, furthermore, to estimate the corresponding relaxation rate. First, we outline the methodology for the simulation of third-order PP spectra of a plexcitonic system.

The plexcitonic system is described within the Tavis–Cummings model.[56] This model describes the interaction of $N$ two-level systems with a single quantized electromagnetic field mode. The corresponding Hamiltonian in the rotating-wave approximation is

$$H = \hbar\omega_c a^\dagger a + \hbar\omega_m \sum_{i=1}^{N} \sigma_i^+ \sigma_i^- + \hbar g \sum_{i=1}^{N} (\sigma_i^+ a + a^\dagger \sigma_i^-) \,, \tag{10}$$

where $a$ ($a^\dagger$) is the annihilation (creation) operator of the electromagnetic field mode, i.e., the SPP, of frequency $\omega_c$, $\sigma_i^-$ ($\sigma_i^+$) is the Pauli lowering (raising) operator of the $i^{\mathrm{th}}$ two-level system of frequency $\omega_m$, and $g$ is the strength of the light–matter coupling. The energy difference between the resonance energy of the two-level system and of the electromagnetic field mode is commonly called detuning. The delocalization length of excitons in thin films is typically in the range of tens of molecules.[57] Compared with the decay lengths of SPPs on a gold surface of a few micrometers, the delocalization lengths of excitons are small. Thus, we assume that there are spatially well separated exciton domains in the thin film such that the excited states of the exciton domains all couple to a common SPP mode but not between different exciton domains. Then, each two-level system in Eq. (10) resembles the ground state and the excited state of a single exciton domain, which contributes to the main exciton transition (Fig. 1f). Thus, we refer to the two-level system as excitons. While the Tavis-Cummings model is not suitable for a microscopic description of the annihilation process due to missing intermolecular couplings, this shortcoming does not limit its applicability to simulate the third-order optical response of the relaxation dynamics following LP excitation. Diagonalization of the Hamiltonian (Eq. 10) results in the eigenstates of the plexcitonic system. For excitons, the transition dipole moment determines if and with which intensity a transition between exciton states is observable in the absorption spectrum. For plexcitons, Lidzey et al. found that intensity of a transition between the ground state and a one-particle plexciton state in a linear absorption measurement is determined by the coefficient of the SPP basis state in the wavefunction of the plexciton state.[58] Based upon this, we use the non-diagonal matrix elements of the operator $a + a^\dagger$, i.e., the photonic transition moments, to determine the intensity of a transition between arbitrary plexciton states. This is different from ordinary molecular spectroscopy, where transition dipole moments determine transition intensities. The different behavior can be rationalized by considering that for plexcitons excited in the Kretschmann configuration, it is a valid approximation that the laser is not transmitted through the 50 nm gold layer such that the electric field strength of the laser vanishes in the adjacent molecular layer. Therefore, not the molecular degrees of freedom are driven but the photonic ones. For transitions between the ground state and a one-particle plexciton state, the matrix elements of the photonic transition moment correspond to the result of Lidzey et al. and they extend to transitions between plexciton states with an arbitrary particle number. Since $a$ and $a^\dagger$ change the particle number by ±1, transitions are only possible between states whose particle number differs by 1 [SI, Section VI, Eq. (S3)]. We will describe the one- and two-particle states of the Hamiltonian and the allowed transitions between those in the following. For a detailed description of the corresponding wavefunctions and the size of the photonic transition moments see SI, Section VI.

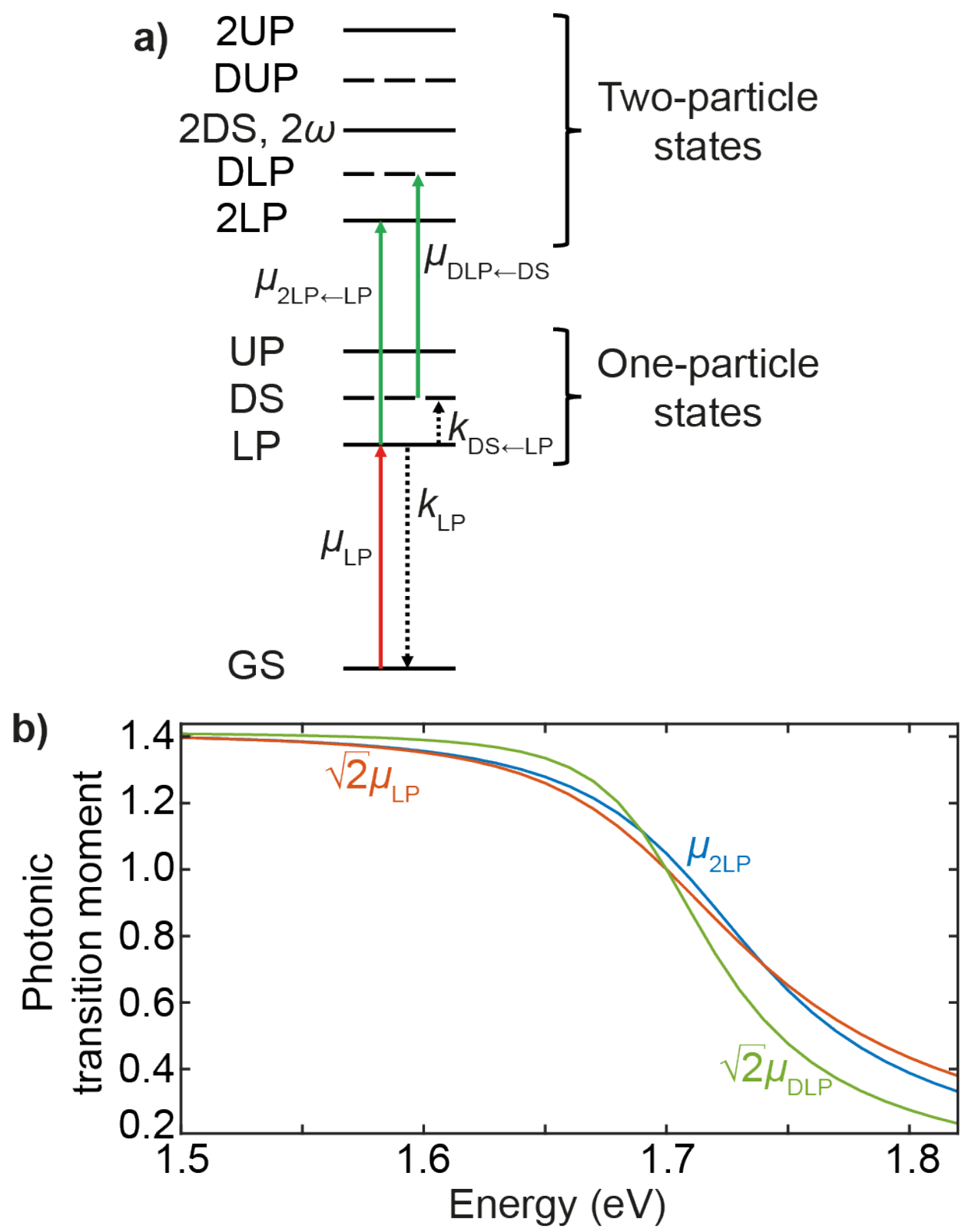

Figure 3. State structure of the Tavis–Cummings model. (a) One- and two-particle states predicted by the Tavis–Cummings model without detuning. Allowed optical transitions are indicated by solid-line arrows and the assumed relaxation processes indicated by dashed arrows. Only those transitions are shown that are included in the modeling. (b) Dependence of the photonic transition moment of the three relevant transitions (see (a)) on the resonance energy of the SPP. The photonic transition moments were calculated within the Tavis–Cummings model with a small number of exciton states ($N = 3$) such that the differences are clearly visible.

The eigenstates of the Tavis–Cummings model without detuning and the possible transitions from the states populated within the dynamics of the LP and DS are depicted in Figure 3a. The red solid line illustrates the excitation from the GS and the green solid lines represent the possible ESA transitions. The dashed lines depict the possible relaxation pathways. The one-particle eigenstates consist of one LP, one UP, and $N - 1$ dark states (DS). While the LP and UP are mixed states with partially photonic and excitonic character, the DS are purely excitonic states. Furthermore, the DS are such linear combinations of the single-exciton states that the photonic transition moment from the GS to any DS vanishes (SI, Section VI), whereas the LP and UP are optically bright from the GS due to their photonic character.

The two-particle eigenstates consist of one 2LP, one $2\omega$ (biplexcitonic state at two times the exciton energy under resonance), one 2UP, $N - 1$ dark lower polaritons (DLP), $N - 1$ dark upper polaritons (DUP), and $\binom{N}{2} - N$ second dark states (2DS). Just like the DS, the 2DS have purely excitonic character and vanishing photonic transition moments from any of the one-particle states, whereas the DLP and DUP states are optically accessible via the DS manifold. The 2LP, $2\omega$, and 2UP states can be considered as biplexcitonic states and have partially photonic and excitonic character. Thus, they have non-vanishing photonic transition moments from either LP or UP or even both. The energies of the 2LP and 2UP symmetrically split up around the 2DS. Although the $2\omega$ state and the 2DS are degenerate if no detuning is present, the two types of states have different character. The 2DS are optically not accessible via any of the one-particle states and have no photonic character. The $2\omega$ state, however, can be transiently excited from LP and UP and has mixed photonic and excitonic character. Furthermore, if detuning is introduced, the energy of the $2\omega$ state changes and is approximately given by the mean value of the 2LP and 2UP energies. Since the energy of the 2DS stays constant as a function of the SPP resonance frequency, the two states are not degenerate anymore if detuning is present.

For the simulation of the third-order PP spectra, we numerically calculated the eigenenergies and photonic transition moments (see Fig. 3b) of the plexcitonic states for a system consisting of $N = 10$ excitons. Although realistically plexcitons can consist of $10^3$ to $10^5$ excitons coupled to a common mode of a SPP[10], $N = 10$ excitons represents a compromise between a computationally treatable and a realistic system. Furthermore, we have checked that our results are consistent if no detuning is present, where analytical expressions for an arbitrary number of excitons are available (SI, Section VI). The resonance energy of the excitons and the strength of the light–matter coupling were taken from literature and set to $\hbar\omega_m = 1.75$ eV and $\hbar g\sqrt{N} = 0.035$ eV, respectively.[47] Since the angle of incidence is linked to the resonance frequency of the excited SPP mode, we fitted the resonance frequency of the SPP mode for the calculations of the static properties such that the crossing point from negative to positive signal in the third-order PP spectra were the same for the experimental PP spectra and the simulated ones. With these parameters we obtained SPP resonance energies at the three different angles 44.1°, 44.5°, and 44.9° to be $\hbar\omega_c = 1.53$ eV, $\hbar\omega_c = 1.70$ eV, and $\hbar\omega_c = 1.78$ eV, respectively. For the three different angles of incidence, all spectra calculated within this model were broadened with a Gaussian distribution of width $\sigma = 0.045\text{ eV} + \Gamma$. Here, $\Gamma$ is the natural linewidth of the LP and the constant value of $0.045$ eV corresponds approximately to the width of the third-order PP spectrum of the ZnPc/glass sample. This broadening probably stems from static or dynamic disorder in the thin film.

Besides the static properties of the one- and two-particle eigenstates of the Tavis–Cummings model, we also simulated the dynamics of the one-particle states to calculate third-order PP spectra of the plexcitonic system. We included in our simulations a simple kinetic model: In this model, the spectrum of the pump pulse only overlaps with the absorption of the LP so that we can define the LP as the initial state. The subsequent kinetic evolution contains two relaxation pathways (Fig. 3a, dashed lines).

First, relaxation occurs from LP to GS with rate constant $k_{\mathrm{LP}}$. This relaxation pathway is induced by the finite lifetimes of the two independent systems, i.e., excitons and SPP. Since the picosecond to nanosecond lifetime of the exciton[55] is long compared to the femtosecond lifetime of the SPP, we neglect the contribution of the exciton lifetime. The corresponding rate can be calculated from the lifetime of the SPP and the wavefunction of the LP,[59]

$$k_{\mathrm{LP}} = \kappa \left| \langle \mathrm{GS} | a + a^{\dagger} | \mathrm{LP} \rangle \right|^2, \tag{11}$$

with the given GS and the SPP relaxation rate $\kappa$. Here, the expectation value of the photonic transition moment operator reduces to the coefficient of the photonic basis state in the LP wavefunction. Thus, the rate $k_{\mathrm{LP}}$ increases with the photonic character of the LP. The lifetime of an SPP is given by $\frac{L}{v_g}$, where $L$ is the propagation length and $v_g$ is the group velocity. For an SPP at an Au/air interface with resonance frequency $\omega$, the dielectric function of Au, $\epsilon(\omega) = \epsilon'(\omega) + i\epsilon''(\omega)$, determines the SPP dispersion $\omega(k)$. The dispersion is the inverse of $k(\omega) = \frac{\omega}{c}\sqrt{\frac{\epsilon'(\omega)}{1+\epsilon'(\omega)}}$ and the group velocity is the derivative of the SPP dispersion, $v_g = \frac{\partial \omega}{\partial k}$. The propagation length of an SPP at an Au/air interface is given by $L = \frac{1}{k}\frac{2\epsilon'(\epsilon'+1)}{\epsilon''}$. We calculated the propagation length and the group velocity of an SPP with a given resonance frequency from Au's dielectric function, which is known from literature.[60] In this way, we obtain SPP lifetimes at the three different incidence angles of 44.1°, 44.5°, and 44.9° to be $\tau = 95$ fs, $\tau = 118$ fs, and $\tau = 173$ fs, respectively.

The second relaxation pathway is the decay from LP to the DS manifold with rate constant $k_{\mathrm{DS}\leftarrow\mathrm{LP}}$. Since it is known from literature that the LP can relax to the DS manifold[16], we add this pathway to our model empirically. As we will explain later, the total relaxation rate of the LP is in our model given by $k_{\mathrm{LP}} + k_{\mathrm{DS}\leftarrow\mathrm{LP}}$ and thus we can estimate $k_{\mathrm{DS}\leftarrow\mathrm{LP}}$ from the experimental total relaxation rate of the LP and the calculated value of $k_{\mathrm{LP}}$.

We neglect relaxation from DS to GS. Since the DS manifold has pure excitonic character, the rate would correspond to the process connected to the largest time constant (10 ns) of the ZnPc/glass sample. Therefore, this process will not affect the short-time dynamics of the system and thus we do not include it in the simulation.

According to these considerations, the total kinetics of the system at early times can be described by a parallel decay of the LP to the DS manifold and to the GS. Analytical solutions for the kinetics of a parallel decay are well known (for details see SI, Section VIII). Thus, it can be shown that the population of the DS and the GS rise with the same rate, which is the sum of the rates of the individual relaxation pathways. If the complete population is initially transferred to the LP, the ratio of the DS population and the GS population is not time-dependent but is given by the ratio of the corresponding rate constants.

While we obtain quantum mechanical eigenstates from the Hamiltonian (Eq. 10), our kinetic model does not capture the quantum nature of the plexciton dynamics. We resort to this propagation scheme because it represents a simple way to include the relaxation process from LP to the DS manifold and to capture the qualitative trends of the measurements for different angles. Furthermore, we only need to calculate population changes of the plexciton states to simulate the PP spectra. The use of the Lindblad equation, which is a quantum mechanical propagation scheme, leads to exponential decay of the populations with the same rates as in the kinetic model. This justifies our approach of using a kinetic model.

With the calculated energies of the eigenstates and the photonic transition moments between the eigenstates, we calculate the third-order PP spectra within a response-function formalism. This is done by following all possible pathways in Liouville space, represented by double-sided Feynman diagrams, and calculating their contributions to the system response (SI, Section VII).

### III.3 Comparing Model and Experiment

The annihilation rate extracted from experiment by global analysis of the fifth-order nonlinear response (Fig. 2c) is not affected by the light–matter interaction between the SPP and the main exciton transition of ZnPc. Therefore, we conclude that the long-time energy transport is also not affected by the coupling of the SPP with the exciton. To understand the unaffected long-time energy transport, we analyze the single sub-picosecond time constant of the third-order PP spectra of the ZnPc/Au sample for every angle of incidence. The population distribution in the involved states helps us understand the long-time dynamics. Note that for the measurement at $\alpha$= 44.1°, this sub-picosecond time constant is on the same time scale as the coherent artifact[61–64] and therefore too small for a reliable fit. Accordingly, we assume the time constant at 44.1° to be given by 65 fs, which we estimated from the length of the coherent artifact and the time constants at 44.5° and 44.9°. The fitted time constants of the other two third-order PP spectra (44.5° and 44.9°) decrease with increasing plasmonic character (decreasing angle of incidence) as expected for the plexcitonic system (SI, Section V and Tab. S2). Therefore, we attribute this time constant to the decay of the LP.

To gain further insight into the decay dynamics of the LP, we now discuss and compare the experiment together with the model simulations from Section III.2. The early-time signals of the experimental third-order PP spectra at the three different incidence angles are shown in Figure 4a–c, together with the simulated spectra in Figs. 4d–f. First, we want to discuss the origin of the ESA contribution in the ZnPc/Au sample which is absent in the ZnPc/glass sample. The coupling of the exciton to the SPP leads to the emergence of new eigenstates in the two-particle manifold at approximately twice the LP resonance energy. The Tavis–Cummings model[56] predicts two transitions from the one- to the two-particle manifold, both of which can possibly lead to ESA: the optically bright transition from the LP to the 2LP and the optically bright transition from the DS manifold to the DLP. The transition energy of both transitions is slightly larger than that of the GS-to-LP transition, which results in a blue shift of the positive ESA signal compared to the GSB/SE of the LP. Since the energy difference of the LP-to-2LP (DS-to-DLP) transition and the GS-to-LP transition is much smaller than the width of the corresponding signals, the signals overlap. In this case, the distance of the maxima of the absolute values of the ESA and the GSB/SE is not determined by the transition energies but by the width of the signals.

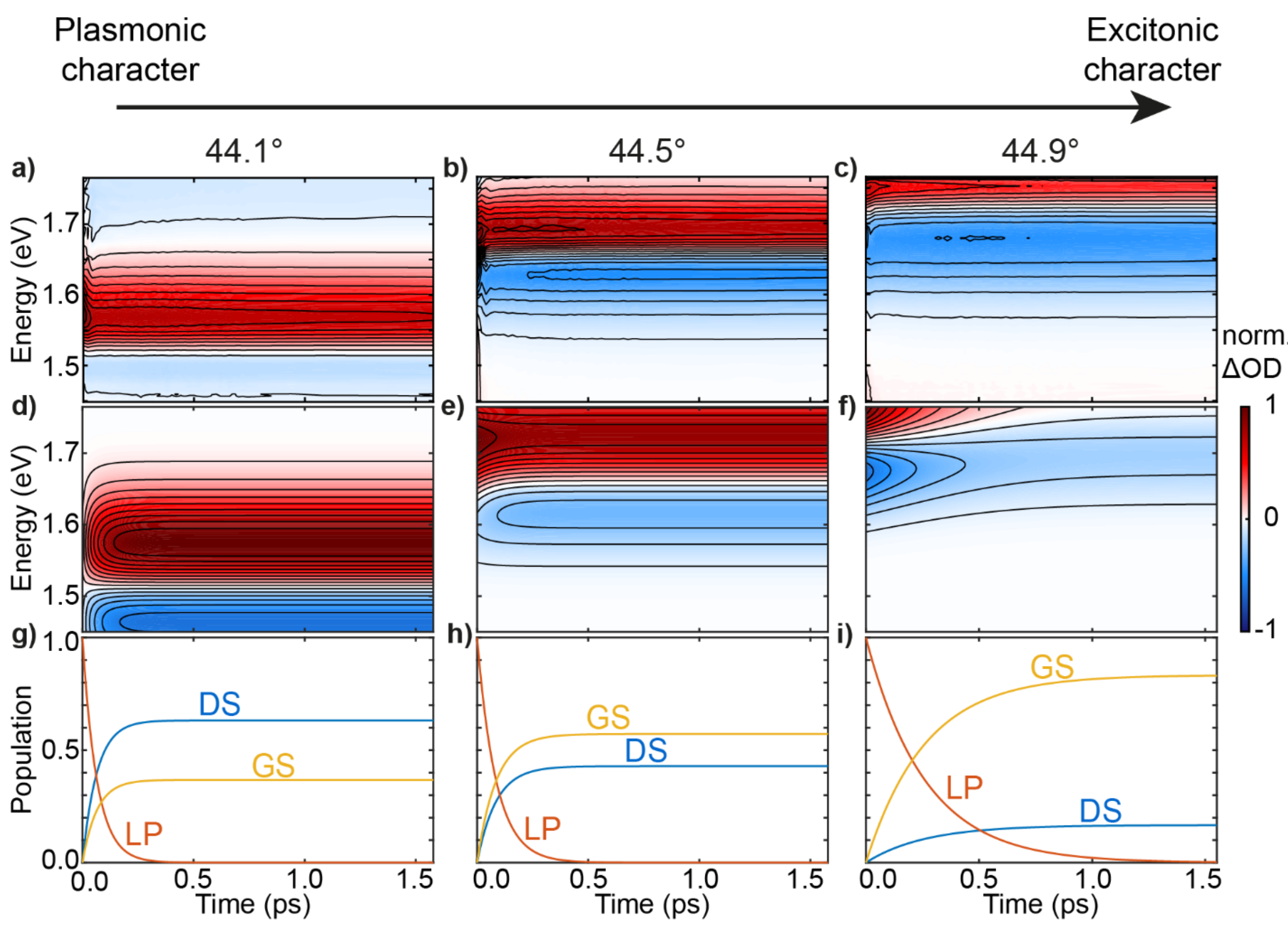

Figure 4. Early-time dynamics of the plexciton in ZnPc/Au for different angles of incidence of the laser beam (44.1°, 44.5°, and 44.9° from left to right). The excitonic character increases from left to right. (a–c) Experimental third-order signals. (d–f) Simulated third-order signals. (g–i) Populations of the lower polariton state (LP), ground state (GS), and dark states (DS) calculated with a kinetic model.

An important finding of the population dynamics at all three different incidence angles is that the overall lifetime of the LP is short (< 300 fs). In all cases, the population of the LP after 1 ps is nearly zero and the only populated excited state is the DS. Since the LP relaxes fast (< 300 fs) to the DS, it is fair to assume that also the 2LP state relaxes to the 2DS on a similar timescale. The lifetime of the 2LP and of the states populated within the two-particle manifold determine the annihilation rate in the fifth-order plexciton spectra. If we compare the lifetime of the 2LP with the timescale of the annihilation rate found in the fifth-order signal (Fig. 2c) of about 146 ps, we see a difference of over two orders of magnitude. This difference means that the vast majority of annihilation events takes place long after the 2LP relaxed to the 2DS. Therefore, the diffusion is dominated by the diffusion of the 2DS population. The 2DS have purely excitonic character and thus we assume that the diffusion properties are excitonic. This agrees with the finding that the annihilation of the pure excitonic ZnPc/glass sample shows the same annihilation rate as the plexcitonic ZnPc/Au sample. In both samples the transport is dominated by a pure excitonic state (the 2DS for the plexcitonic system and the excitonic state for the excitonic system).

Note that the dark state (DS) is called dark because it cannot be directly excited from the GS with light. Nevertheless, ESA is possible as described above. In addition, if the DS is populated the GS is less populated (Fig. 4g–i). This leads to a bleach of the GS and therefore to a GSB signal. So, although the DS is "dark" from the GS, ESA and GSB signals can be detected in PP spectroscopy.

The third-order signal at $\alpha$ = 44.5° shows an interesting feature in the measurement (Fig. 4b) and in the simulations (Fig. 4e). The positive signal decays over the population time while the negative signal rises. If the dynamics of the plexciton were only determined by the decay of the LP to the GS, there would be no increase in the absolute value of any of the signals. Therefore, a second decay channel must play a key role in the dynamics of the LP to explain the observed signal rise. The simulated spectra with $k_{\mathrm{DS}\leftarrow\mathrm{LP}} = 0$ and the populations of the involved states (GS, LP, DS) are shown in SI, Section IX. The decay from the LP to the DS leads to two modifications in the optical response of the system. First, if the LP relaxes into the DS manifold, SE from the LP can no longer take place which leads to a decrease of the absolute value of the negative signal contribution. Second, the ESA is altered since the photonic transition moment between the DS manifold and the DLP manifold differs from the photonic transition moment between the LP and the 2LP. If the number of molecules coupled to the electromagnetic field mode were infinite and if there were no detuning between the frequencies of the electromagnetic field and the molecules, the ratio of the two photonic transition moments would be given by $\frac{\mu_{2LP\leftarrow LP}}{\mu_{DLP\leftarrow DS}} = \sqrt{2}$. Since the contribution of a given Feynman diagram to the PP spectrum, which we will call for example SE signal for the SE Feynman diagrams, is proportional to the absolute square of the photonic transition moments, the change in the ESA and the SE signals as a function of time induced by the relaxation from the LP to the DS would exactly cancel each other out. However, in a system in which the number of molecules is finite, this is not the case. For an incidence angle of 44.5° (SPP energy of 1.7 eV), the photonic transition moment between LP and 2LP is slightly larger than $\sqrt{2}\mu_{LP}$ (Fig. 3b) and thus, the ESA decreases relatively to the SE as a function of time, which leads to an increase in the absolute value of the negative signal and a decrease in the positive signal. The situation is similar with an angle of incidence of 44.9° (Figs. 4c and f).

At 44.1°, the simulations predict an increase of absolute magnitude of the negative spectral contribution as well as of the positive ESA contribution (Fig. 4d). If the resonance energy of the SPP lies far below the exitonic resonance energy, the energy difference of the LP and the 2LP state is much closer to the energy difference of the LP to the GS than the energy difference of the DS to the DLP. Therefore, the signals connected to the LP (ESA from LP to 2LP and GSB+SE of the LP) are smaller than the signals connected to the DS manifold (GSB and ESA from DS to DLP), i.e., the overall signal increases if the system relaxes into the DS manifold. The measured signal (Fig. 4a) does not show this behavior so clearly, as our time resolution is not sufficient to resolve this phenomenon.

In addition to the PP spectra, we also show the population of the GS, LP and DS manifold in Figs. 4g–i. Comparing the populations at different incidence angles, it can be seen that the ratio of the GS population to the population of the DS manifold increases with the incidence angle, i.e., the ratio of the decay rates $k_{\mathrm{LP}}$ and $k_{\mathrm{DS}\leftarrow\mathrm{LP}}$ increases. $k_{\mathrm{LP}}$ and $k_{\mathrm{DS}\leftarrow\mathrm{LP}}$ become smaller with increasing excitonic character. Since the change of $k_{\mathrm{LP}}$ is smaller than the change of $k_{\mathrm{DS}\leftarrow\mathrm{LP}}$, the ratio of $k_{\mathrm{LP}}$ and $k_{\mathrm{DS}\leftarrow\mathrm{LP}}$ increases. Since $k_{\mathrm{DS}\leftarrow\mathrm{LP}}$ decreases with increasing angle of incidence, we conclude that the larger the contribution of the SPP to the LP, the larger is the relaxation rate. Therefore, we speculate that the relaxation from the LP to the DS is mainly caused by the finite lifetime of the SPP, which is in accordance with recent theoretical studies.[17,18]

## IV. Conclusion

We used higher-order PP spectroscopy to investigate the excitation energy transport of a plexcitonic system consisting of a thin film of ZnPc on an SAM-covered Au substrate. We measured third- and fifth-order PP spectra of the sample via higher-order PP spectroscopy.[42,44] For this method, a linear combination of different pump intensities is used to separate the different nonlinear orders. By varying the incidence angle $\alpha$, the properties of the coupled LP can be tuned. We obtained third- and fifth-order PP spectra for three different angles $\alpha$ and thus varied the plasmonic/excitonic character of the LP state. We compared the plexciton signals with the third- and fifth-order signals of a ZnPc/glass sample, which exhibits purely excitonic character.

We found that our PP signal shifts in energy as a function of the incidence angle $\alpha$, like the linear spectra. Therefore, we observed signals with different ratios of excitonic and plasmonic contributions to the plexcitonic wave function. However, we found that the long-time dynamics of the plexcitonic sample and more importantly the time constant of the annihilation are almost independent of the angle of incidence and resemble that of the excitonic ZnPc/glass sample. From these findings we conclude that the long-time energy transport and lifetime do not depend on the SPP. To explain these observations, we analyzed the early-time dynamics of the third-order signals and compared them with simulations based on the Tavis–Cummings model, as done by Scholes et al on a comparable sample structure.[25]

As proposed in literature, the observation that the annihilation is independent of the angle of incidence led us to the conclusion that the LP can quickly relax into the dark-states (DS) manifold besides the relaxation to the GS.[25,65,66] We showed that this assumption leads to a correct description of the angle-dependent third-order PP spectra for small delay times. Furthermore, it enabled us to estimate the relaxation rate from the LP to the DS manifold from the experimental decay rate of the LP and the relaxation rate from the LP to the GS.

We conclude that the LP relaxes quickly into the DS. This provides insight into why the long-time transport in the plexcitonic system remains unaffected by the angle of incidence and is nearly identical to the transport in a purely excitonic system. This can be attributed to the fact that the transport occurs over times that are significantly longer than the decay time of the LP. Consequently, the diffusion process is primarily governed by the properties of the DS. Since the DS exhibit a purely excitonic character, the diffusion within the system also retains this purely excitonic nature. As a result, the ZnPc-inherent diffusion process leading to annihilation is independent of the surface plasmon polariton (SPP) and, therefore, does not rely on the angle of incidence or the coupling between the SPP and the exciton.

In literature, enhanced energy transport of polaritons compared to pure excitons has been reported across various samples.[8,10,11,37] These findings seem to contradict the results in this paper, but we believe that the relationship between energy transport and the coupling of excitons with a light field is very complex. There are several underlying factors why some studies report enhanced energy transport while others do not. One reason could be that we detect the annihilation signal of our sample. Due to the short lifetime of the LP, the transport might only differ in the first hundred femtoseconds. For example, Berghuis et al. observed enhanced transport only in the first 100 fs after excitation (when the LP state is populated).[10] Nevertheless, we do not observe an increased number of annihilation events during the first hundred femtoseconds. The annihilation behavior of plexcitons is not fully understood and needs further investigation. Possible explanations for the lack of an enhanced annihilation in the early times could be because the LP states are "protected" against annihilation due to e.g., high delocalization or that the energy transport in the LP is not significantly faster than the transport in the excitonic states. Another reason why some studies report enhanced energy transport while others do not could be the light mode distribution of the studied polariton. Gonzalez-Ballestero et al. report that the DS can have a delocalized character if the electromagnetic field has a discrete spectrum.[67] Therefore, in polariton samples with discrete light modes, such as molecules in a microcavity, energy transport could be enhanced even if LP-to-DS scattering occurs. Lastly, the reason and mechanism of LP-to-DS scattering, as well as the properties of the DS itself, are not fully understood. For example, Parolin et al. report a quasi-dark state that has an oscillator strength unequal to zero.[48] Such unexpected properties and the mechanism of the LP-to-DS scattering that is not fully understood make it difficult to predict the influence of the DS in our sample.

Coupling excitons with light modes has proven to be an effective method for enhancing transport properties. To explore these transport properties, we employed higher-order PP spectroscopy, which can investigate transport via quasiparticle interactions, marking the first application of this technique on plexcitonic systems. Transport measurements in such systems are not possible using conventional third-order spectroscopy such as transient absorption or coherent two-dimensional electronic spectroscopy. Our results indicate that transport to the DS manifold poses a significant challenge in improving transport properties. In the future we will investigate if the role of DS in the context of transport might be changed with larger coupling between the light mode and the exciton. A stronger coupling leads to a greater energy difference between the LP and the DS. Therefore, the scattering between these states could be affected. Another approach is to use a substrate with discrete modes, e.g., by using an optical cavity. In such systems the DS are not purely exitonic but rather delocalized like the light mode.[67] The difference between the annihilation of polaritons, such as plexcitons, and that in pure molecular systems is a compelling scientific question that requires further theoretical investigation and experimental data to be fully understood.

## Supplementary Material

See supplementary material for details on plexciton sample preparation, Kretschmann setup for linear measurements, higher-order pump–probe spectroscopy, pump polarization dependency, global analysis of third- and fifth-order pump–probe spectra, eigenstates of the Tavis–Cummings Hamiltonian, response function formalism for third-order transient absorption spectra, parallel decay kinetics in the plexcitonic sample, simulations without LP-to-DS relaxation, possible errors for the simulations, and calculation of the plexciton dispersion.

## Acknowledgement

T.B. acknowledges funding by the European Research Council (ERC) within Advanced Grant IMPACTS (No. 101141366). L.N.P. acknowledges a fellowship by the FCI. All authors acknowledge financial support by the Bavarian State Ministry for Science and the Arts within the collaborative research network “Solar Technologies go Hybrid” (SolTech).

## Conflict of Interest

The authors declare no competing interest.

## Author Contributions

S.B. and L.N.P. contributed equally to this work.

**Simon Büttner**: Formal analysis (equal) Investigation (equal) Visualization (lead) Writing - original draft (equal). **Luca Nils Philipp**: Formal analysis (equal) Investigation (equal) Writing - original draft (equal). **Julian Lüttig**: Formal analysis (supporting) Writing - original draft (equal) Writing - review & editing (equal). **Maximilian Rödel**: Formal analysis (supporting) Resources (lead). **Matthias Hensen**: Formal analysis (supporting) Supervision (equal) Writing - review & editing (equal). **Jens Pflaum**: Conceptualization (equal) Formal analysis (supporting) Funding acquisition (equal) Project administration (equal) Supervision (equal) Writing - review & editing (equal). **Roland Mitric**: Conceptualization (equal) Formal analysis (supporting) Funding acquisition (equal) Project administration (equal) Supervision (equal) Writing - review & editing (equal). **Tobias Brixner**: Conceptualization (equal) Formal analysis (supporting) Funding acquisition (equal) Project administration (equal) Supervision (equal) Writing - review & editing (equal).

## Data Availability

The data that support the findings of this study are openly available in Zenodo at https://doi.org/10.5281/zenodo.15262172.[68]

# Supporting Information

# Probing plexciton dynamics with higher-order spectroscopy

Simon Büttner[1#], Luca Nils Philipp[1#], Julian Lüttig[1,2], Maximilian Rödel[3], Matthias Hensen[1], Jens Pflaum[3,4‡], Roland Mitric[1+], and Tobias Brixner[1,5*]

*[1]Institut für Physikalische und Theoretische Chemie, Universität Würzburg, Am Hubland, 97074 Würzburg, Germany*
*[2]Department of Physics, University of Michigan, 450 Church Street, Ann Arbor, Michigan 48109, USA*
*[3]Experimental Physics VI, University of Würzburg, Am Hubland, 97074 Würzburg, Germany*
*[4]Center for Applied Energy Research e.V. (CAE Bayern), Magdalene-Schoch-Straße 3, 97074 Würzburg, Germany*
*[5]Center for Nanosystems Chemistry (CNC), Universität Würzburg, Theodor-Boveri-Weg, 97074 Würzburg, Germany*

*[#]Contributed equally to this work*

*[‡]Electronic mail:* jens.pflaum@uni-wuerzburg.de
*[+]Electronic mail:* roland.mitric@uni-wuerzburg.de
**Electronic mail:* tobias.brixner@uni-wuerzburg.de

## I. Plexciton Sample Preparation

For the sample preparation, we used 1 mm thick standard object holders which were cleaned in an ultrasonic bath for 15 min immersed in a water/Mucasol, an acetone and then an isopropanol solution. After that, we thermally evaporated a 2 nm thick chromium wetting layer followed by a 50 nm thick gold layer in a vacuum chamber with a base pressure of $10^{-6}$ mbar. As an anti-quenching layer, a self-assembled monolayer (SAM) of 1-decanethiol was chemisorbed on top of the gold thin film. Finally, 10 nm of the purified organic semiconductor zinc phthalocyanine (ZnPc) was prepared via molecular beam deposition under high vacuum with a base pressure of $10^{-8}$ mbar.

## II. Kretschmann Setup for Linear Measurements

The excitation of the plexciton dispersion was performed via a Kretschmann setup.[1] For that, we used a halogen lamp white-light source (SLS201L/M, Thorlabs) with a collimator package (SLS201C). For the p-polarized light, necessary for excitation of the coupled exciton–plasmon state, we used two polarizing beamsplitters (CCM1-PBS251/M and CM05-PBS202, Thorlabs) to cover a wide wavelength range of 420 nm–1000 nm. The used half-ball lens is manufactured out of N-BK7 glass by hand (J. HAUSER GmbH

& Co. KG) and has an additional grind down at the plane side to compensate the thickness of the sample and to ensure the perfect alignment of the axis of rotation. As a refractive matching, we used a low-auto-fluorescent oil (IMMOIL-F30CC, Olympus) to prevent refractive effects at the sample-half ball interface. The wavelength-dependent detection of the light was conducted by a spectrometer (MAYA2000Pro, Ocean Optics) equipped with an S11510-1106 detector chip (Hamamatsu), an H13 grating and a fixed slit with a width of 10 µm. The rotation of the excitation arm and the sample was performed with two rotational stages.

## III. Higher-Order Pump–Probe (PP) Spectroscopy

Higher-order pump–probe (PP) spectroscopy is a nonlinear optical technique that allows for the differentiation between single-particle dynamics and multi-particle interactions. In traditional PP measurements, contributions from various nonlinear orders are combined, making it difficult to distinguish between different excitation phenomena. The higher-order PP spectroscopy method employs intensity-dependent measurements to systematically isolate nonlinear responses.[2–4]

For the different incident angles ($\alpha$) of the ZnPc/Au sample, we used three different intensities to extract nonlinear responses up to the seventh-order response. For these measurements we used pump pulses as described in Section II in the main manuscript {pulse duration of ~19 fs [intensity full width at half maximum (FWHM)]; beam profile of an elliptical shape with a short axis length of about 70 µm (intensity FWHM) and a long axis length of about 130 µm (FWHM)} with a base intensity of $I_0$= 1.5 nJ (and according to Eq.2 in the main manuscript $I_2$= 4.5 nJ and $I_3$= 6.0 nJ). We chose the intensity so that the seventh-order signal was smaller than the noise to ensure that signals from even higher orders than seventh order are negligible. To check if the extraction is correct, we measured also at a very low intensity of $I_{\text{ref}}$= 0.25 nJ and scaled this signal to the extracted third-order signal. The same procedure was used for the ZnPc/glass sample. For the pure molecular sample, we used the same pump pulse as described before with a base-intensity of $I_0$= 12.5 nJ (and according to Eq.2 in the main manuscript $I_2$= 37.5 nJ and $I_3$= 50.0 nJ). The low-intensity reference intensity was set to $I_{\text{ref}}$= 2.0 nJ.

Table S1. Intensities used for the higher-order PP spectroscopy measurements.

| | $I_3$= 4$I_0$ | $I_2$= 3$I_0$ | $I_0$ | $I_{\text{ref}}$ |
|---|---|---|---|---|
| 44.1° | 6.0 nJ | 4.5 nJ | 1.5 nJ | 0.25 nJ |
| 44.5° | 6.0 nJ | 4.5 nJ | 1.5 nJ | 0.25 nJ |
| 44.9° | 6.0 nJ | 4.5 nJ | 1.5 nJ | 0.25 nJ |
| Glass | 50 nJ | 37.5 nJ | 12.5 nJ | 2.0 nJ |

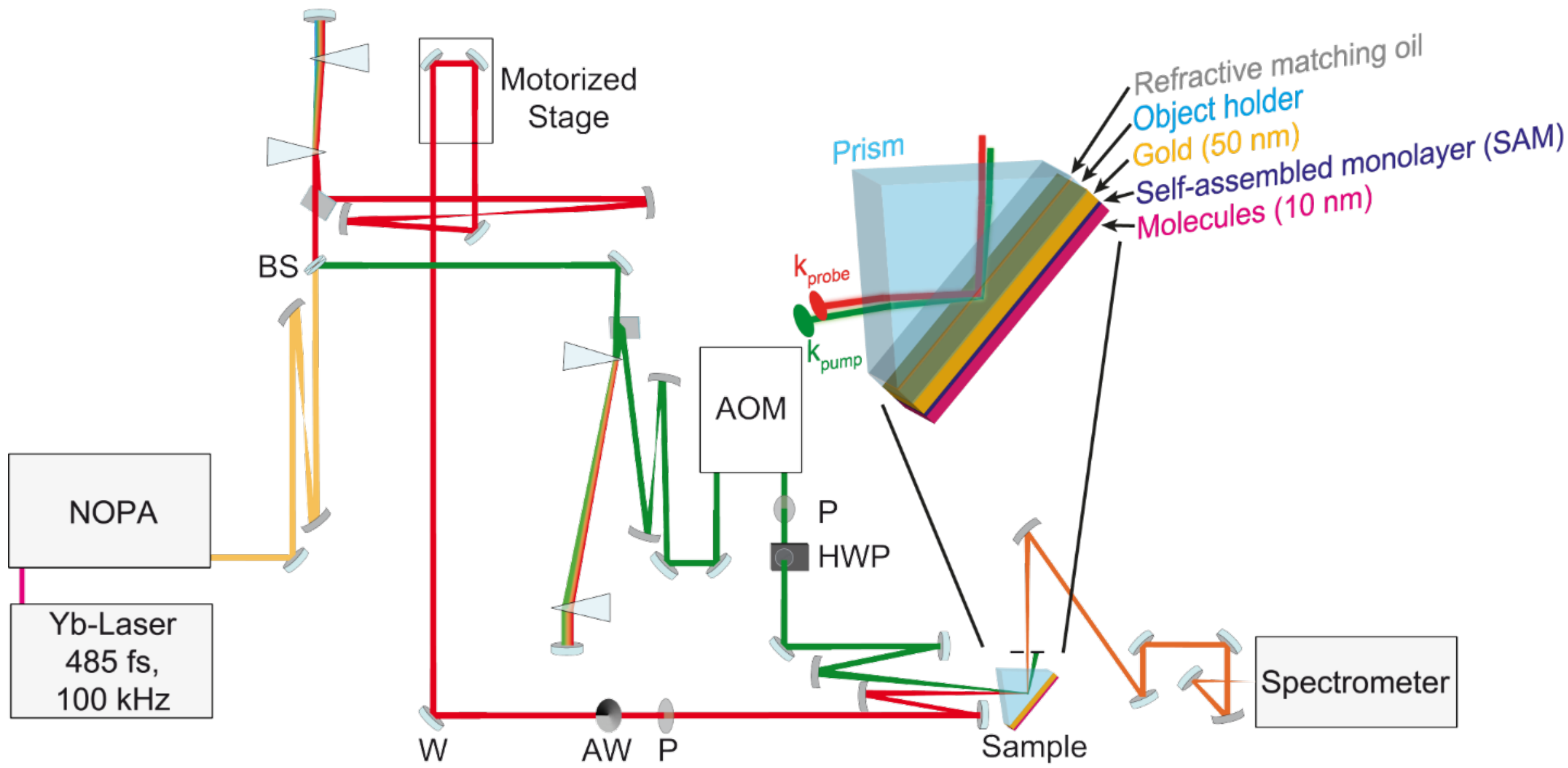

Figure S1. Schmatic overview of the experimental setup. To separate the pump (green) and probe (red) beams a beam splitter (BS) was utilized. For the pump pulse a wedged window (W), an attenuation weel (AW) and a polarizer (P) was used. For the pump beam an acousto-optic modulator (AOM) pulse shaper, a polarizer (P) and a half-wave plate (HWP) were applied.

The setup was described in the main text. A schematic is shown in Figure S1.

## IV. Pump Polarization Dependency

The excitation of a pure molecular excitonic system, such as our polycrystalline ZnPc film, is less dependent on the laser polarization as an SPP system. By coupling the excitonic and plasmonic systems, the newly formed plexciton has the same polarization dependency as the SPP. Therefore, the plexciton excitation is much more sensitive to the laser polarization, than the pure exciton. This polarization dependency can be used to check if the observed PP signal arises from a pure excitonic system or from a coupled plexcitonic system.

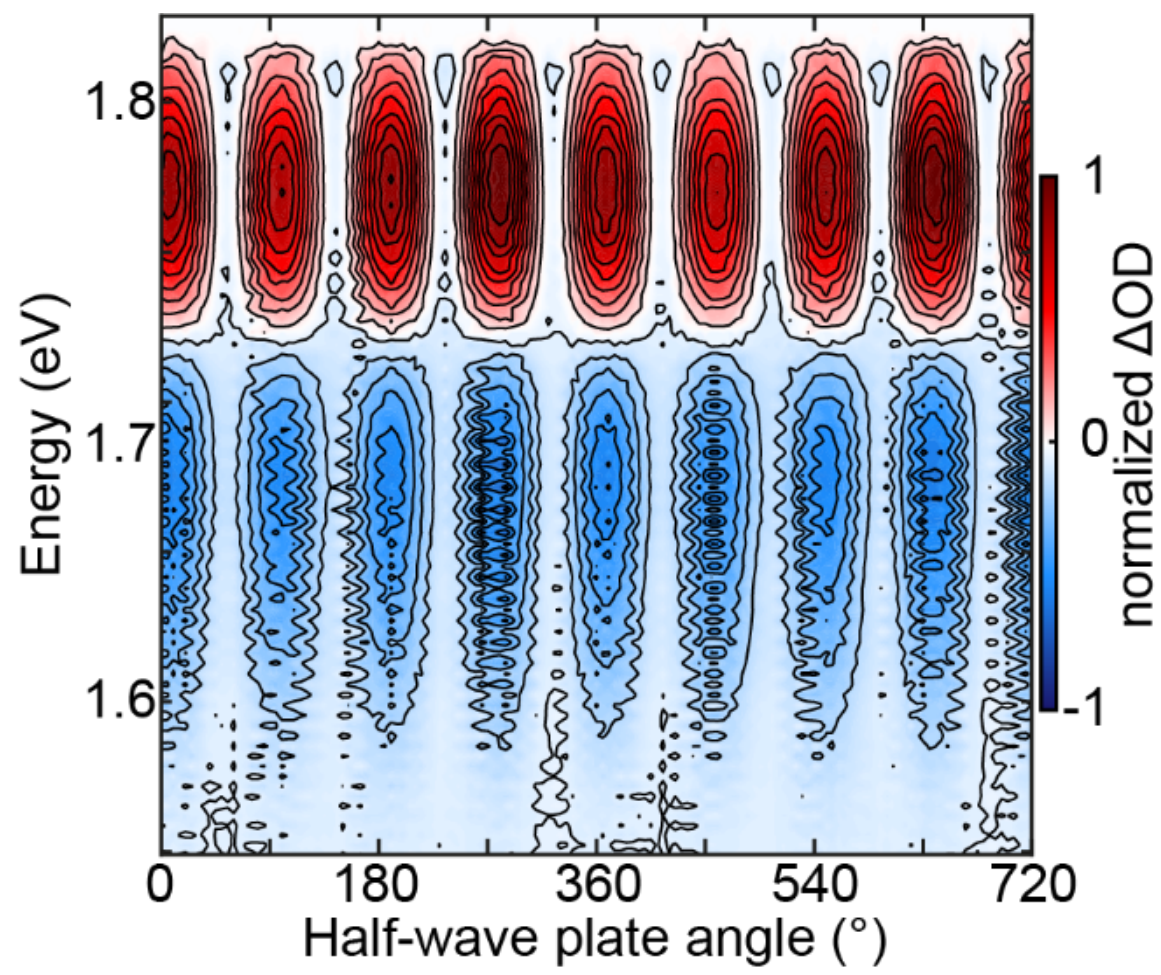

Figure S2. Pump–probe (PP) signal of the ZnPc/Au sample at an incidence angle $\alpha$= 44.5° at a population time of 500 fs for different half-wave plate angles.

We measured the PP signal of the ZnPc/Au sample at an incidence angle $\alpha$= 44.5° at a fixed population time of 500 fs for different half-wave plate angles for the pump pulse (Fig. S2). The probe pulse polarization was kept parallel. The observed signal has an oscillation period of 90° which correspond to a switch from parallel to perpendicular polarization. At the minimum the signal is very close to zero. The positive as well as the negative signal are dependent on the pump laser polarization. This is a strong indicator that the PP signal arises from a state that is connected to an SPP. Together with the observed shift along the detection axis with the incident angle, we conclude that the observed signals in the ZnPc/Au sample are coupled plexciton signals and not pure molecular signals.

## V. Global Analysis of Third- and Fifth-Order Pump-Probe (PP) Spectra

To extract the time constants of the lifetime and the annihilation rate, we used global analysis of the third- and fifth-order pump-probe (PP) spectra. The global analysis was carried out with Glotaran 1.5.1.[5] We fitted the third-order spectra of the different measurement geometries and samples with three (44.1°) or four (44.5°, 44.9°) time constants ($\tau_1$–$\tau_4$). Furthermore, we used an instrument response function (IRF) with one gaussian function and a third-order polynomial dispersion curve to fit the time zero and the pulse overlap signal of each measurement. The same IRF and dispersion curve was used to fit the fifth-order spectrum at the same sample geometry. An additional time constant ($\tau_{\text{Annihilation}}$) was added for the fifth-order spectrum and the time constants $\tau_1$–$\tau_4$ were set fixed to the same values as the time constants of the third-order spectrum. The decay-associated spectra (DAS) are plotted in Figure S3, and the values of the time constants are listed in Table S2.

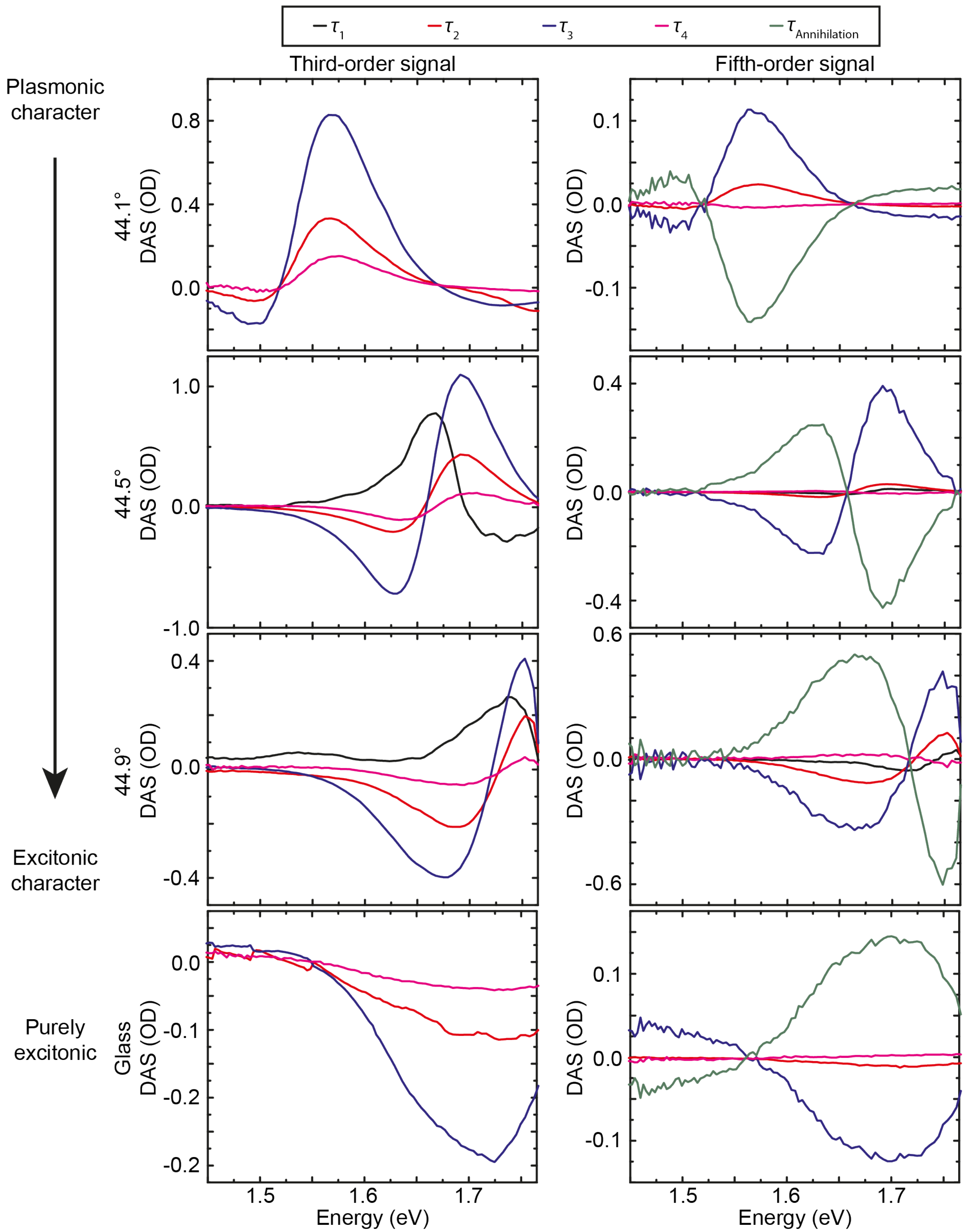

Figure S3. Decay-associated spectra (DAS) from the global analysis of the third-order (left) and fifth-order spectra (rigth) for ZnPc/Au at incidence angles of 44.1° (first row), 44.5° (second row), and 44.9° (third row). In the fourth row, the DAS from the global analysis of the third-order (left) and fifth-order spectra (right) for ZnPc/glass are shown.

Table S2. Time constants extracted by global analysis for measured third- and fifth-order PP spectra

| | Third and fifth order | | | | Fifth order |
|---|---|---|---|---|---|
| | $\tau_1$ (ps) | $\tau_2$ (ps) | $\tau_3$ (ps) | $\tau_4$ (ns) | $\tau_{\text{Annihilation}}$ (ps) |
| 44.1° | – | 6.6 ± 0.24 | 119 ± 1.9 | 10 | 147 ± 3.8 |
| 44.5° | 0.085 ± 0.0021 | 7.3 ± 0.21 | 129 ± 1.4 | 10 | 140 ± 2.6 |
| 44.9° | 0.26 ± 0.0067 | 7.4 ± 0.24 | 129 ± 1.8 | 10 | 162 ± 6.4 |
| Glass | – | 2.6 ± 0.21 | 112 ± 5.1 | 10 | 136 ± 6.8 |

The time constant $\tau_4$ is too long to measure it accurately with our setup due to the maximum delay time of 6 ns.[6] Therefore, the time constant was fixed at 10 ns for all measurements. The fastest time constant $\tau_1$ for the 44.1° measurements is too short to be isolated from the IRF of the setup and the so-called coherent artefact.[7,8] For the measurement of ZnPc on glass we only used three time constants ($\tau_2$–$\tau_4$). The second time constant $\tau_2$ is smaller for the ZnPc/glass sample compared to the ZnPc/Au measurements. This could be caused by the different substrate and the sample preparation or because the coupling of the exciton with the surface-plasmon polariton (SPP) stabilizes the plexciton quasiparticle resulting in a larger time constant.

## VI. Eigenstates of the Tavis–Cummings Hamiltonian

In this section we provide analytical solutions for the one- and two-particle eigenstates of the Tavis-Cummings Hamiltonian without detuning. The Tavis–Cummings Hamiltonian describes the interaction between an ensemble of $N$ two-level systems coupled to a single electromagnetic field mode. If there is no detuning, the corresponding Hamiltonian is given by

$$H = \hbar\omega a^\dagger a + \hbar\omega \sum_{i=1}^{N} \sigma_i^+ \sigma_i^- + \hbar g \sum_{i=1}^{N} (\sigma_i^+ a + a^\dagger \sigma_i^-), \tag{S1}$$

where $a$ ($a^\dagger$) is the annihilation (creation) operator of the electromagnetic field mode of frequency $\omega$, $\sigma_i^-$ ($\sigma_i^+$) is the Pauli lowering (raising) operator of the $i^{\text{th}}$ two-level system of frequency $\omega$ and $g$ is the strength of the light–matter coupling. This Hamiltonian commutes with the generalized number operator $N_p = a^\dagger a + \sum_{i=1}^{N} \sigma_i^+ \sigma_i^-$. Thus, the eigenstates of the Hamiltonian are also eigenstates of the generalized number operator. Eigenstates to $N_p$ with eigenvalue 1 (2) are often referred to as the one- (two-)particle eigenstates.

Usually, the Tavis–Cummings Hamiltonian is described in the bare-state basis, i.e., the basis obtained by considering a direct product of the eigenstates of the constituting systems. One-particle states within this basis are given by $|G,1\rangle$ and $|0,e_i\rangle$. $|G,1\rangle$ can be interpreted as the state in which the ensemble of two-level systems is in its ground state, while the electromagnetic field mode is in its first excited state. Similarly, $|0,e_i\rangle$ is the state in which the electromagnetic field mode is in its ground state, while the $i^{\text{th}}$ two-level system is in its excited state and all other two-level systems in the ground state.

### VI.1 One-Particle Eigenstates

The one-particle eigenstates of the Tavis–Cummings Hamiltonian consist of the lower polariton (LP), upper polariton (UP), and $N-1$ dark states. The corresponding eigenstates are given by

$$|\mathrm{UP/LP}\rangle = \frac{1}{\sqrt{2}}|G,1\rangle \pm \frac{1}{\sqrt{2N}}\sum_{i=1}^{N}|e_i,0\rangle, \tag{S2}$$

with eigenenergies of $E_{\mathrm{UP/LP}} = \hbar\omega \pm \hbar g\sqrt{N}$. The photonic transition moment of these eigenstates is given by $\langle G,0|a+a^{\dagger}|\mathrm{UP/LP}\rangle = \frac{1}{\sqrt{2}}$.

The dark states are purely molecular states at an energy of $E_{\mathrm{DS}} = \hbar\omega$, which can be written as

$$|\mathrm{DS}\rangle = \sum_{i=1}^{N} c_i|e_i,0\rangle, \tag{S3}$$

where the coefficients have to fulfill $\sum_{i=1}^{N} c_i = 0$. Thus, the photonic transition moment from the ground state to any of the DS vanishes.

### VI.2 Two-Particle Eigenstates

The two-particle basis states within the bare basis are given by $|G,2\rangle$, $|e_i,1\rangle$, and $|e_ie_j,0\rangle$. All two-particle eigenstates of the Tavis–Cummings model can be expressed in terms of these basis states. The energetically highest and lowest eigenstates are the second lower (2LP) and upper (2UP) polaritons, respectively, which can be written as

$$|\mathrm{2UP/2LP}\rangle = \frac{1}{\sqrt{2}}\sqrt{\frac{N}{2N-1}}|G,2\rangle \pm \frac{1}{\sqrt{2N}}\sum_{i=1}^{N}|e_i,1\rangle + \frac{1}{\sqrt{N(2N-1)}}\sum_{i=1}^{N}\sum_{j>i}^{N}|e_ie_j,0\rangle \tag{S4}$$

with eigenenergies $E_{\text{2UP/2LP}} = 2\hbar\omega \pm 2\hbar g\sqrt{N - \frac{1}{2}}$. The 2UP (2LP) can be transiently excited from the UP (LP) state. The corresponding photonic transition moment is equal for both transitions, $\langle \text{UP}|a + a^\dagger|\text{2UP}\rangle = \langle \text{LP}|a + a^\dagger|\text{2LP}\rangle = \frac{1}{2} + \sqrt{\frac{N}{2(2N-1)}}$.

Furthermore, there is a third state contained in the two-particle eigenstates, which can be transiently excited from LP and UP, the so-called 2ω state, which is given by

$$|2\omega\rangle = \sqrt{\frac{N-1}{2N-1}}|G, 2\rangle - \sqrt{\frac{2}{(N-1)(2N-1)}}\sum_{i=1}^{N}\sum_{j>i}^{N}|e_i e_j, 0\rangle. \quad \text{(S5)}$$

Energetically, this state is located at twice the energy of the electromagnetic field mode $E_{2\omega} = 2\omega$ and the photonic transition moment from LP and UP is $\langle \text{LP}|a + a^\dagger|2\omega\rangle = \langle \text{UP}|a + a^\dagger|2\omega\rangle = \sqrt{\frac{N-1}{2N-1}}$.

Also, the model predicts two manifolds of states between the 2UP (2LP) and the 2ω state, which are commonly referred to as the DUP (DLP) states. Just like the DS among the one-particle eigenstates, DUP and DLP manifolds consist of $N - 1$ states and can be written as

$$|\text{DUP/DLP}\rangle = \sum_{i=1}^{N} c_i|e_i, 1\rangle \pm \frac{1}{\sqrt{N-2}}\sum_{i=1}^{N}\sum_{j>i}^{N}(c_i + c_j)|e_i e_j, 0\rangle, \quad \text{(S6)}$$

where the coefficients must fulfill $\sum_{i=1}^{N} c_i = 0$. The energy of these states is given by $E_{\text{DUP/DLP}} = 2\hbar\omega \pm \hbar g\sqrt{N-2}$. DUPs and DLPs cannot be excited from UP nor LP, however, both are optically accessible from the DS manifold. If one chooses an orthonormal representation of the DS and uses the same set of coefficients $c_i$ to parameterize the DUP and DLP wavefunctions, then exactly one DLP and DUP is bright from the corresponding DS with the same coefficients with photonic transition amplitude $\langle \text{DS}|a + a^\dagger|\text{DLP}\rangle = \langle \text{DS}|a + a^\dagger|\text{DUP}\rangle = \frac{1}{\sqrt{2}}$. While this photonic transition amplitude equals the photonic transition amplitude from the GS-to-UP and GS-to-LP transitions under resonance conditions, this is not true if the energy of the electromagnetic field mode is detuned from the transition of the two-level systems.

Finally, there is a collection of states at the energy of twice the energy of the electromagnetic field mode, which are optically dark from any of the one-particle eigenstates, the so-called second dark states (2DS). The wavefunction of these states can be written as

$$|\text{2DS}\rangle = \sum_{i=1}^{N}\sum_{j>i}^{N} c_{ij}|e_i e_j, 0\rangle, \quad \text{(S7)}$$

where the coefficients must satisfy $\sum_{j\neq i}^{N} c_{ij} = 0$.

## VII. Response Function Formalism for Third-Order TA Spectra

When the interaction between the laser pulses and the system is treated perturbatively, the third-order transient absorption (TA) signal arises due to two interactions of the system with the pump pulse, followed by one interaction with the weak probe pulse.[9] While we assume that the two interactions with the pump pulse happen simultaneously, the interaction with the probe pulse is delayed by the waiting time $T$ with respect to the interactions with the pump pulse. The third-order response function is given by

$$S^{(3)}(t,T) = \left(-\frac{i}{\hbar}\right)^3 \mathrm{Tr}\left(\mu U(t)\left[\mu, U(T)\left[\mu,\left[\mu,\rho_{\mathrm{eq}}\right]\right]\right]\right), \tag{S8}$$

where $\mu$ is the corresponding transition moment and $U$ is the propagator of the system, which will be described in detail in Section VIII. For plexcitons, we use the photonic transition moment $\mu = a + a^{\dagger}$. The response function contains various terms that correspond to different ways how the system can interact with the light pulses. These so-called Liouville-space pathways are conveniently depicted by double-sided Feynman diagrams, which are a pictorial representation of the evolution of the system density matrix in Liouville space.[7,10] The double-sided Feynman diagrams, which are needed for the calculation of the third-order TA signal if the pump and probe pulses are resonant with the LP only, are shown in Figure S4. We here consider strict time ordering between the pump and the probe interactions and consider only diagrams in the phase-matching direction of $-\mathbf{k}_{\mathrm{pump}} + \mathbf{k}_{\mathrm{pump}} + -\mathbf{k}_{\mathrm{probe}}$ . Furthermore, we restrict our analysis to processes that involve population transfer neglecting any coherences between electronic states during the population time. Note that we only show the so-called rephasing diagrams where the first interaction occurs with $-\mathbf{k}_{\mathrm{pump}}$. There is an equivalent set of diagrams (non-rephasing) for which the two pump interactions are switched and the first interaction occurs with $+\mathbf{k}_{\mathrm{pump}}$. However, the two sets of diagrams only differ in the coherence between the two pump pulse interactions which is not resolved in ordinary TA spectroscopy. When reading those diagrams, a few rules must be followed: Time always runs from bottom to top, bold arrows indicate interactions with the laser pulses, dotted arrows mark the emission of the spectroscopic signal, and dotted vertical lines represent the system dynamics within the waiting time $T$ between the pump and the probe pulse. Due to the commutator in the expression of the response function, the contribution of a given Feynman diagram must be multiplied by a factor of $(-1)^n$, where $n$ is the number of interactions with the laser fields from the right. As an example, for how a specific Liouville-space pathway contributes to the response function, the term corresponding to the stimulated emission Feynman diagram is given by

$$S_{SE}^{(3)}(t,T) = \left(-\frac{i}{\hbar}\right)^3 \mu_{LP}^2 U_{LP,LP}(T)\mu_{LP}^2 \mathrm{e}^{-i\frac{E_{LP}t}{\hbar}}, \tag{S9}$$

where $U_{LP,LP}(T)$ is the probability that if the system is initially in the LP, it also is in the LP after the waiting time $T$. Furthermore, we have ignored relaxation within the last time interval $t$. All other pathways contribute analogously. The total response function is then obtained by summing over all contributing Liouville space pathways,

$$S^{(3)}(t,T) = \sum_{i=1}^{5} S_i^{(3)}(t,T). \tag{S10}$$

In the semi-impulsive limit, the total response equals the third-order polarization[7,9]

$$S^{(3)}(t,T) = P^{(3)}(t,T). \tag{S11}$$

Finally, the third-order TA spectrum $PP^{(3)}(\omega,T)$ is proportional to the imaginary part of the Fourier-transformed third-order polarization[7,9]

$$PP^{(3)}(\omega,T) \propto -2\mathrm{Im}\left[P^{(3)}(\omega,T)\right]. \tag{S12}$$

Like the third-order response function, the third-order TA spectrum consists of contributions by the different Liouville-space pathways. For example, the contribution of the stimulated emission Feynman diagram is given by

$$PP_{SE}^{(3)}(\omega,T) \propto -\mu_{LP}^2 U_{LP,LP}(T)\mu_{LP}^2 \delta\left(\omega - \frac{E_{LP}}{\hbar}\right). \tag{S13}$$

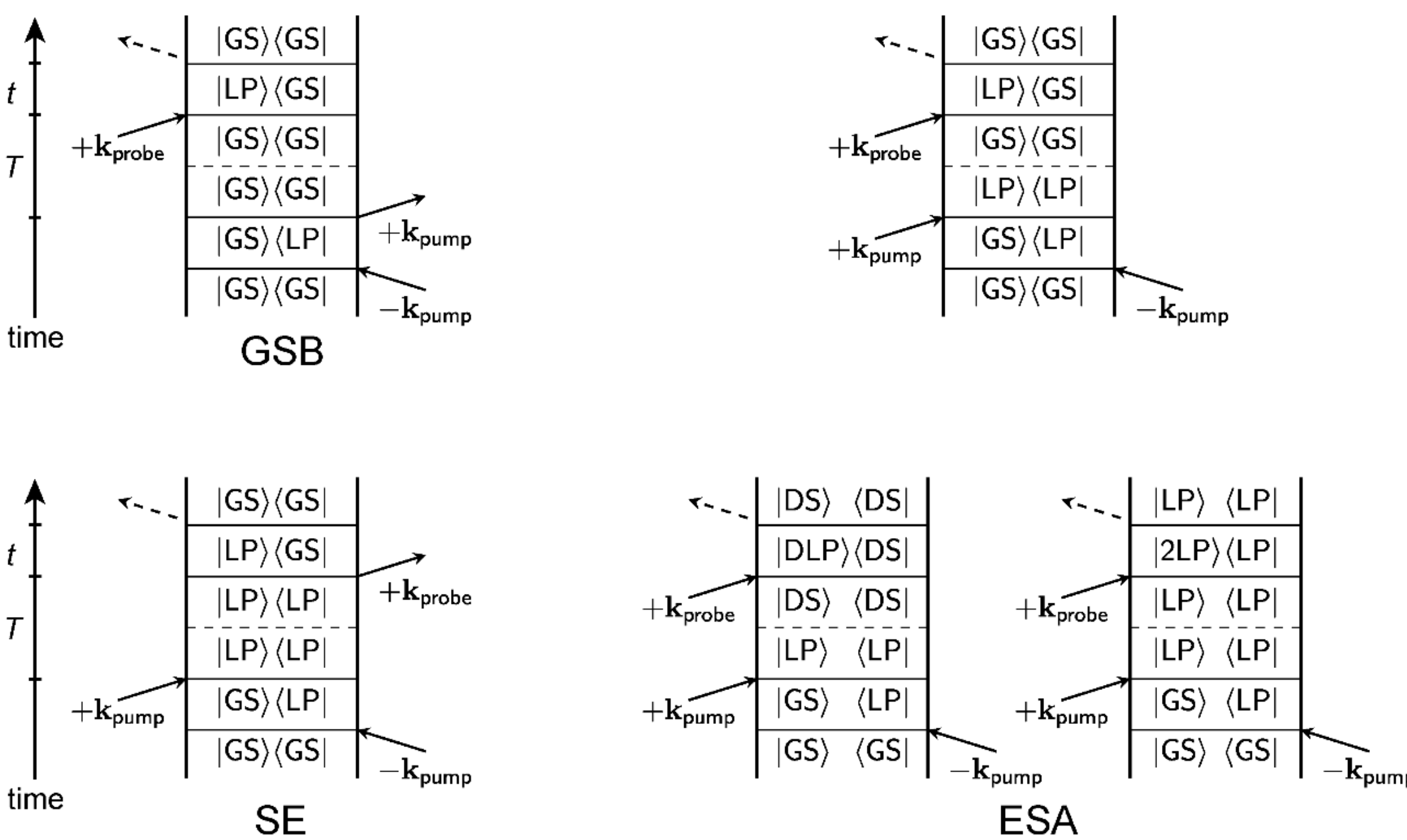

Figure S4. Double-sided Feynman diagrams depicting the different Liouville-space pathways that contribute to the third-order TA signal.

## VIII. Parallel Decay Kinetics in the Plexcitonic Sample

When encountering a parallel decay, the population of a single state, in our case the population of the LP, $p_{\mathrm{LP}}$, relaxes simultaneously into two different states, in our case the GS and the DS with populations $p_{\mathrm{GS}}$ and $p_{\mathrm{DS}}$, respectively. In this case, the time-dependent populations of all three states are described by a system of coupled differential equations,

$$\frac{\mathrm{d}p_{\mathrm{LP}}}{\mathrm{d}t} = -k_1 p_{\mathrm{LP}} - k_2 p_{\mathrm{LP}}, \tag{S14}$$

$$\frac{\mathrm{d}p_{\mathrm{GS}}}{\mathrm{d}t} = k_1 p_{\mathrm{LP}}, \tag{S15}$$

$$\frac{\mathrm{d}p_{\mathrm{DS}}}{\mathrm{d}t} = k_2 p_{\mathrm{LP}}, \tag{S16}$$

with the two rate constants $k_1$ and $k_2$. These equations are easily solved, which yields the time-dependent populations of the three states,

$$p_{\mathrm{LP}}(t) = p_{\mathrm{LP}}(0)e^{-(k_1+k_2)t}, \tag{S17}$$

$$p_{\mathrm{GS}}(t) = \frac{k_1 p_{\mathrm{LP}}(0)}{k_1+k_2}\left(1 - e^{-(k_1+k_2)t}\right), \tag{S18}$$

$$p_{\mathrm{DS}}(t) = \frac{k_2 p_{\mathrm{LP}}(0)}{k_1+k_2}\left(1 - e^{-(k_1+k_2)t}\right), \tag{S19}$$

where we assumed that the GS and DS are not populated at $t = 0$.

## IX. Simulations without LP-to-DS Relaxation

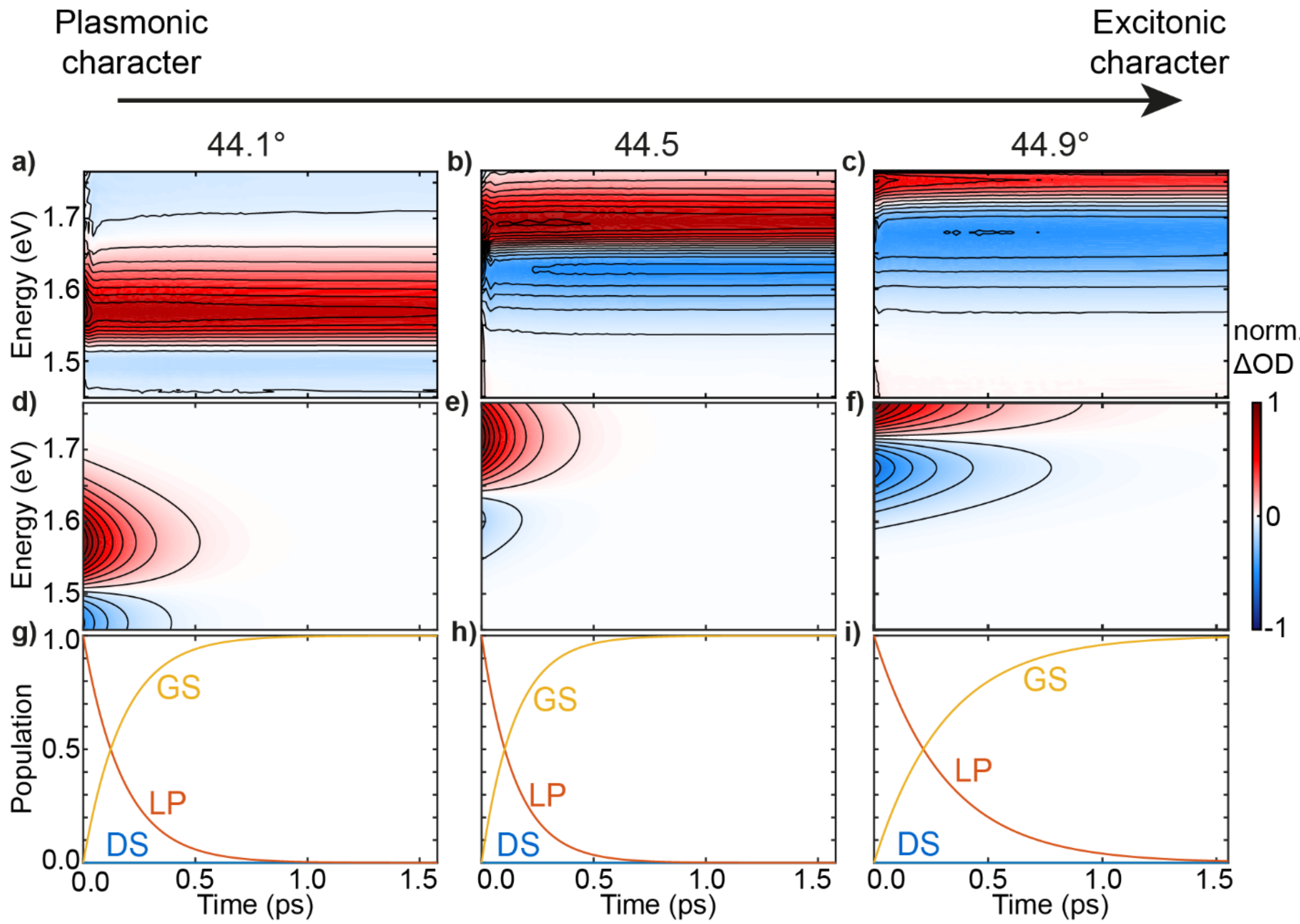

Figure S5. Early-time dynamics of the plexciton in ZnPc/Au for different angles of incidence of the laser beam (44.1°, 44.5°, and 44.9° from left to right). Note that in this simulation, the relaxation from LP to DS is assumed to vanish, $k_{(DS \leftarrow LP)} = 0$. The excitonic character increases from left to right. (a–c) Experimental third-order signals. (d–f) Simulated third-order signals. (g–i) Populations of the lower polariton state (LP), upper polariton state (UP), and dark states (DS) calculated with a kinetic model.

## X. Possible Errors of the Simulations

At an angle of 44.9° the simulated third-order TA spectra show major deviations from the experimental spectra. The biggest difference is that only the positive signal decreases exponentially, while the absolute value of the negative signal stays almost constant. This indicates that the simulations overestimate the ratio $k_{0 \leftarrow \mathrm{LP}}/k_{\mathrm{DS} \leftarrow \mathrm{LP}}$ so that the relaxation of the LP into the ground state is too strong. Since $k_{\mathrm{DS} \leftarrow \mathrm{LP}}$ is calculated from the experimental lifetime of the LP and $k_{0 \leftarrow \mathrm{LP}}$, this can be traced back to an overestimation of $k_{0 \leftarrow \mathrm{LP}}$. This rate is determined by two quantities: the plexcitonic wave function and the lifetime of the SPP. However, those two quantities are also connected, since both depend on the resonance frequency of the SPP. Certainly, one source of error is the determination of the SPP's resonance frequency, which we fit from the experimental spectra as explained in Section VI.1. A second source of error is the calculation of the lifetimes of the SPP modes. We obtain these from the dieletric function of Au and calculate the lifetime for an Au/vacuum interface. However, we do not consider that the ZnPc thin film changes the dielectric surrounding at the interface, which potentially leads to a change in the SPP lifetime.

## X. Calculation of the Plexciton Dispersion

The dispersion spectra of the ZnPc/Au plexcitons were calculated with the so-called coupled oscillator model.[11] Within this model, the single-particle states of the plexcitonic system are obtained by eigenvalue decomposition of the following matrix,

$$\begin{pmatrix} \hbar\omega(k) & V_1 & V_2 \\ V_1 & \hbar\omega_1 & 0 \\ V_2 & 0 & \hbar\omega_2 \end{pmatrix}, \tag{S20}$$

where $\omega_1$ and $\omega_2$ are the resonance frequencies of the two main exciton transitions of the ZnPc thin film, $\omega(k)$ is the dispersion relation of SPP on the gold surface depending on the wavevector $k$, and $V_1$ and $V_2$ are the coupling strengths between the two main exciton transitions and the SPP. From literature it is known that the resonance energies of the two main exciton transitions of the ZnPc thin film and their respective couplings to the SPP are given by $\hbar\omega_1 = 1.75$ eV, $\hbar\omega_2 = 1.97$ eV, $V_1 = 35$ meV, and $V_2 = 47$ meV.[1] The inverse of the dispersion relation of the SPP depends on the real part of the dielectric function of gold, $\epsilon_1{}'(\omega)$, and that of the ZnPc thin film, $\epsilon_2{}'$, which we assume to be constant, leading to

$$k(\omega) = \frac{\omega}{c}\sqrt{\frac{\epsilon_1'(\omega)\epsilon_2'}{\epsilon_1'(\omega)+\epsilon_2'}}. \tag{S21}$$

The values of the dielectric function of gold were taken from literature.[12] Since the penetration depth of the SPP into the ZnPc thin film is larger than the width of the thin film, the dielectric constant of the ZnPc entering equation Eq. S21 is not that of bulk ZnPc. As an approximation, we have taken $\epsilon_2' = 1.05$. Finally, we need to define the visibility of a plexcitonic state in the dispersion spectra. To this end, we followed Lindzey et al. and defined the visibility of a plexcitonic state as the absolute square of the expansion coefficient of the SPP basis state in the eigenvectors of Eq. S20.[13] Using these formulas, we calculated the resonance frequencies, i.e., eigenvalues of Eq. S20, and their visibilities with wave vectors between 7 and 11 $\mu\mathrm{m}^{-1}$. However, these values depend on the wave vector and not on the incidence angle $\alpha$. Thus, they were transformed to depend on incidence angle via

$$\alpha = \arcsin\left(\frac{k\lambda}{2\pi n_{pr}}\right), \tag{S22}$$

where $\lambda$ is the wavelength of the respective plexciton resonance and $n_{pr}$ is the refractive index of the prism used in the Kretschmann geometry (in our case $n_{pr} = 1.5168$). Plotting the visibility depending on the resonance frequency of the plexciton states and the incidence angle and broadening all signals with two-dimensional gaussian distributions of widths $\sigma_1 = 0.2°$ and $\sigma_2 = 0.04$ eV yields Figure 1d of the main paper.

For the calculation of the plexciton dispersion, we have used the coupled oscillator to include the second exciton transition of the ZnPc thin film, which leads to a better description of the UP at high incidence angles. It should be noted that the LP is already well described without inclusion of the second exciton transition. Furthermore, the coupled oscillator model is equivalent to the Tavis–Cummings model for the calculation of the resonance frequencies and visibilities of plexcitonic states. However, this is not the case for plexciton states in higher particle subspaces, where one must use the Tavis–Cummings model to describe those states.